\documentclass[fleqn,usenatbib]{mnras}

\usepackage{newtxtext,newtxmath}

\usepackage[T1]{fontenc}
\usepackage{ae,aecompl}
\usepackage{hyperref}

\usepackage{graphicx}	
\usepackage{amsmath}	
\usepackage{amssymb}	

\title[space.ml: PSFGAN]{\texttt{PSFGAN}: a generative adversarial network system for separating quasar point sources and host galaxy light}

\author[Stark et al.]
{Dominic Stark,$^{1}$\thanks{E-mail: dostark@student.ethz.ch}Barthelemy Launet,$^{1,2}$
Kevin Schawinski,$^{1}$
Ce Zhang,$^{3}$ 
\newauthor
Michael Koss,$^{1,4}$ 
M. Dennis Turp,$^{1}$
Lia F. Sartori,$^{1}$ 
Hantian Zhang,$^{3}$
\newauthor
Yiru Chen,$^{5}$
and Anna K. Weigel$^{1}$
\\
$^{1}$Institute for Particle Physics and Astrophysics, ETH Zurich, Wolfgang-Pauli-Strasse 27, CH-8093, Z\"{u}rich, Switzerland\\
$^{2}$LERMA, Observatoire de Paris, PSL Research Univ., CNRS, Sorbonne Univ., UPMC Univ. Paris 06, 75014 Paris, France\\
$^{3}$Systems Group, ETH Zurich, Universit\"{a}tstrasse 6, CH-8006, Z\"{u}rich, Switzerland\\
$^{4}$Eureka Scientific, 2452 Delmer Street Suite 100, Oakland, CA 94602-3017, USA\\
$^{5}$Institute of Network Computing and Information Systems, Peking University, Beijing, China
}

\date{Accepted XXX. Received YYY; in original form ZZZ}

\pubyear{2017}

\begin{document}
\label{firstpage}
\pagerange{\pageref{firstpage}--\pageref{lastpage}}
\maketitle

\begin{abstract}
The study of unobscured active galactic nuclei (AGN) and quasars depends on the reliable decomposition of the light from the AGN point source and the extended host galaxy light.  The problem is typically approached using parametric fitting routines using separate models for the host galaxy and the point spread function (PSF). We present a new approach using a Generative Adversarial Network (GAN) trained on galaxy images. We test the method using Sloan Digital Sky Survey (SDSS) \emph{r}-band images with artificial AGN point sources added which are then removed using the GAN and with parametric methods using \texttt{GALFIT}. When the AGN point source PS is more than twice as bright as the host galaxy, we find that our method, \texttt{PSFGAN}, can recover PS and host galaxy magnitudes with smaller systematic error and a lower average scatter ($49\%$). \texttt{PSFGAN} is more tolerant to poor knowledge of the PSF than parametric methods. Our tests show that \texttt{PSFGAN} is robust against a broadening in the PSF width of $\pm 50\%$ if it is trained on multiple PSF's. We demonstrate that while a matched training set does improve performance, we can still subtract point sources using a \texttt{PSFGAN} trained on non-astronomical images. While initial training is computationally expensive, evaluating \texttt{PSFGAN} on data is more than $40$ times faster than \texttt{GALFIT} fitting two components. Finally, \texttt{PSFGAN} it is more robust and easy to use than parametric methods as it requires no input parameters.
\end{abstract}

\begin{keywords}
methods: data analysis -- techniques: image processing --  quasars: general
\end{keywords}





\section{Introduction}\label{sec:introduction}

Active Galactic Nuclei (AGN) are among the brightest continuously emitting objects in the Universe radiating in most wavelengths of light.  The link between AGN and the host galaxy properties such as stellar mass \citep[e.g.,][]{stellarmass1, stellarmass2, stellarmass3, stellarmass4} and star formation rate \citep[e.g.,][]{schawinski06, sfr, sfr3, sfr4} are critical to better understand the relationship between black hole growth and the host galaxy. These quantities are frequently inferred by modeling the Spectral Energy Distribution (SED) from multi-wavelength data \citep{simmons2,sed_modeling_4,collinson1,sed_modeling3}. Unfortunately, especially in unobscured quasars, the light from the AGN far outshines the host galaxy emission. Investigating correlations between galaxy parameters and properties of the AGN thus requires a separate analysis of  AGN and host galaxy components \citep{gabor,bias1}.

Extending photometric studies to host galaxies at higher redshift \citep[e.g.][]{bias2,simmons} is critical to understanding their evolution across cosmic time. However for imaging data, if the host galaxy is very faint compared to the quasar and its angular size is close to the width of the Point Spread Function (PSF), it can be hard to detect the host galaxy at all \citep[e.g.,][]{bahcall97}. Following the pioneering work of \cite{bahcall95} the first studies of quasar hosts were conducted using the \textit{Hubble} Space Telescope (HST) \citep[e.g.,][]{mcleod,kirhakos99,hst3,lehnert}. The most widely used techniques were based on scaling and aligning a stellar PSF to the peak of the surface brightness distribution of the quasar. Other approaches included some constraints on the residual host galaxy emission such as monotonicity of the radial light profile \citep{boyce_disney}. These methods however systematically overestimate the quasar contribution and only yield a lower limit for the host galaxy flux. Later studies showed that fitting two-dimensional galaxy components simultaneously with the point source (PS) component yields the most robust method \citep{peng,merger_remnants}. 

One of the most popular methods used for two-dimensional surface profile fitting is \texttt{GALFIT} \citep{peng,peng2}. Its ability to recover PS fluxes and host galaxy parameters has been demonstrated several times both for HST images \citep{simmons,kim,gabor,bias1} and for ground-based images \citep{goulding, mike}. \texttt{GALFIT} is a very powerful tool for detailed morphological decomposition of single cases but it was not designed for batch-fitting \citep{peng}. In the era of Big Data astronomy\footnote{Currently, the total data volume of SDSS is $>125$ TB \citep{sdssdr14}. The LSST will produce $15$ TB of data per year\citep{lsst}.},  where large datasets have to be efficiently analysed without human interaction,  parametric fitting might not be an efficient approach. Nevertheless there have been approaches \citep{galapagos,pymorph} to automate \texttt{GALFIT}  by combining it with \texttt{Source Extractor} \citep{sextractor}, but these methods still depend on their input parameters.  

Machine Learning (ML) often accomplishes the demand for automation and scalability in data analysis. Various ML techniques have been applied to astronomy, for example in outlier-detection \citep{dayla}, galaxy classification \citep{dielemann,galaxy_classification2} or detector characterization \citep{gwave_ml}. The most recent developments in automated galaxy fitting use Bayesian inference \citep{galphat, profit} or deep learning \citep{deeplegato}.

By using a Generative Adversarial Network \citep{GAN} we develop the first ML-based method for separating AGN from their host galaxies. We adopt the \texttt{GalaxyGAN} algorithm \citep{galaxygan} which was originally conceived to recover features in noisy ground-based imaging data. Our method is called \texttt{PSFGAN} as it subtracts point sources from CCD images. We test the effectiveness of \texttt{PSFGAN} at recovering the AGN (and the host galaxy) and compare our results to \texttt{GALFIT}. In section \ref{sec:method} we describe the overall method, we describe the specific GAN architecture in \ref{sec:architecture}, the training and testing procedure in \ref{sec:datasets}, the model selection in \ref{sec:GANmodels} and in \ref{sec:fitting_strategy} the \texttt{GALFIT} fitting strategy we used for the comparisons. In section \ref{sec:results} we test the performance of \texttt{PSFGAN}. Finally, in section \ref{sec:discussion} we discuss applications and limitations.

Throughout this paper, we adopt a cosmology with $\Omega_m=0.3$, $\Omega_\Lambda=0.7$, and $H_0=70$\,km\,s$^{-1}$\,Mpc$^{-1}$.



\section{Method}\label{sec:method}
\subsection{GAN architecture}
\label{sec:architecture}
\begin{figure}
\centering
\includegraphics[width=\columnwidth]{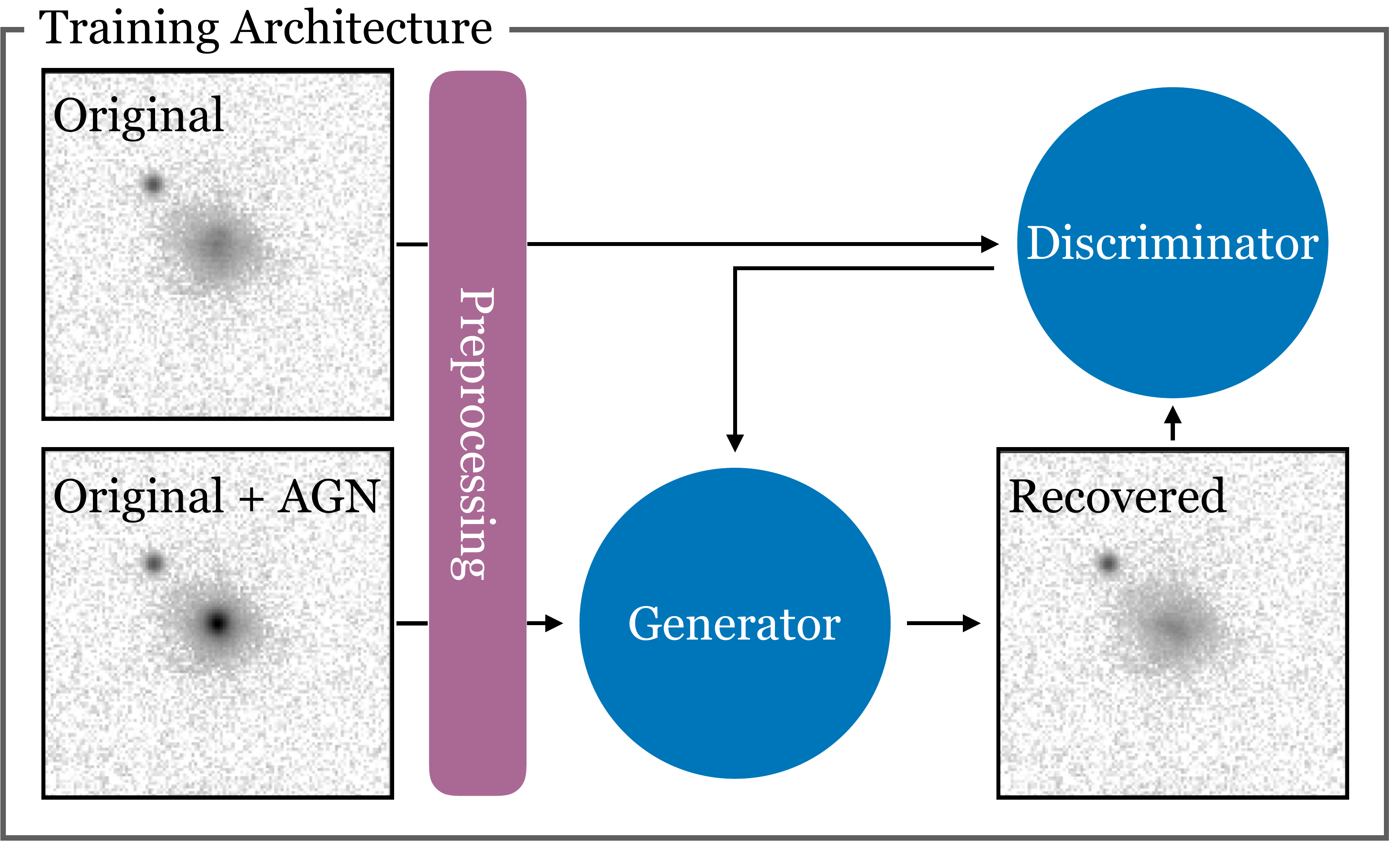}
\caption{Scheme of the architecture used in this work. The generator takes as input the modified image (original galaxy image with a simulated PS in its center) and tries to recover the original galaxy image. The discriminator distinguishes the original from the recovered image. Before feeding the images to the GAN they are normalized to have values in $[0,1]$ and transformed by an invertible stretch function.}
\label{fig:architecture}
\end{figure}

In Figure \ref{fig:architecture}, we show a graphical scheme of the architecture we used.  A GAN consists of two neural networks: a generator and a discriminator. The generator creates artificial datasets, and the discriminator classifies a given set as ``real" or ``fake". The generator and the discriminator are simultaneously trained. In an ideal case, the generator recovers the training data distribution \citep{GAN}. Conditional GANs take a conditional input \citep{condGAN} and can be used for image-processing \citep{pix2pix}. \texttt{GalaxyGAN} takes a degraded galaxy image as conditional input \citep{galaxygan}. During the training the generator tries to recover the original image from the degraded one. The discriminator learns to distinguish between the original image and the generator output. Both networks are trained at the same time to maximize the others loss and by this means the generator learns the inverse of the transformation that has been applied to the original image. In the testing phase, the generator is applied to degraded images it has never seen before, in order to recover the original ones.
In this work we choose the processed image to be the original galaxy image with a simulated PS representing an unobscured AGN. Using this as the conditional input, the generator then learns the inverse transformation which is equivalent to subtracting the PS.

Adding a simulated PS to the center of a galaxy image will primarily affect a few pixels at the center of the image.  We therefore adapt the generator to increase the weight of the central region in the loss computation.
\begin{figure*}
\centering
\includegraphics[width=\textwidth]{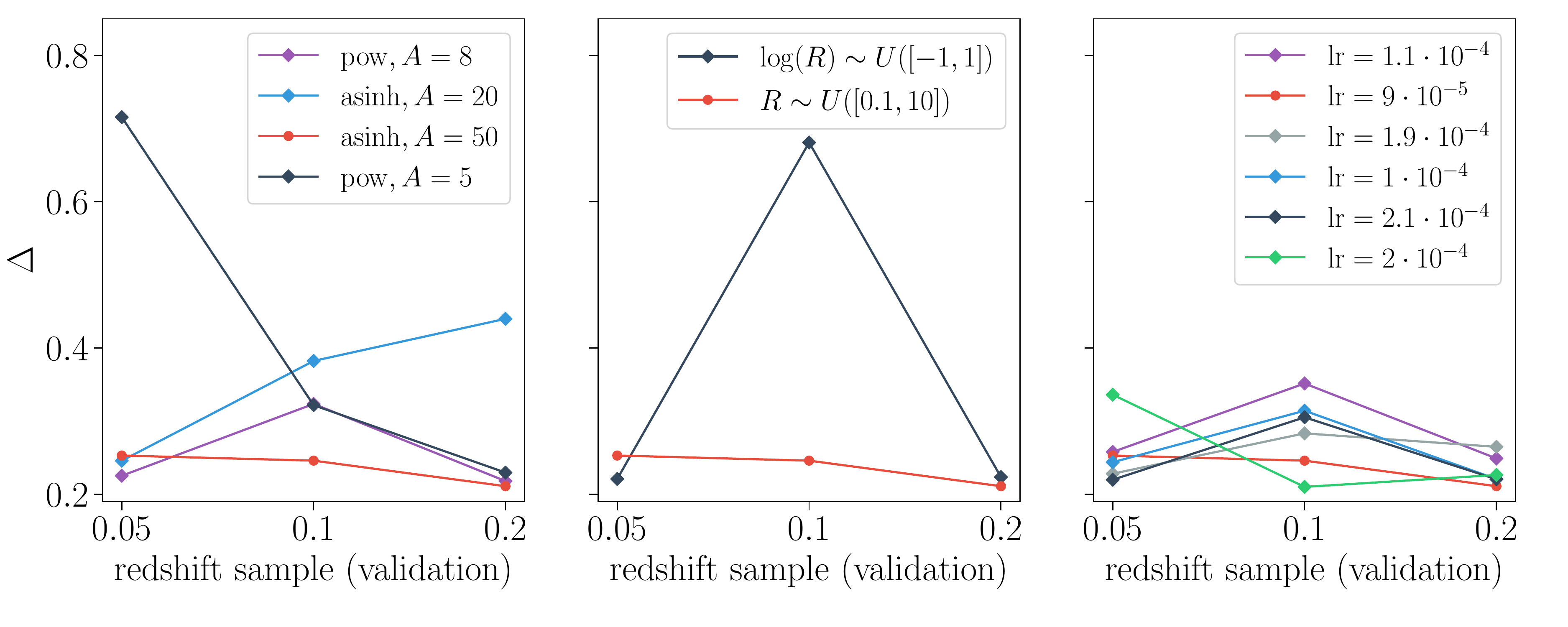}
\caption{The GAN trained with different hyperparameters evaluated on the validation sets of the three redshift samples. We only search a subspace of the whole parameters space. While varying the stretch function we use $\mathrm{lr}=9\cdot 10^{-5}$ for the learning rate and a uniform distribution in linear space for the contrast ratios in the training set. Results for $\log$ and $\mathrm{sigmoid}$ functions are not plotted as their score was more than $5$ times the average score plotted here. While varying the distribution of contrast ratios we use $\mathrm{asinh}, A=50$ as stretch function and $\mathrm{lr}=9\cdot 10^{-5}$ for the learning rate. While varying the learning rates we use $\mathrm{asinh}, A=50$ as stretch function and a uniform distribution in linear space for the contrast ratios. The quantity on the $y$-axis is the MAD of the recovery ratio (recovered PS flux / real PS flux) from $1$.}
\label{fig:MAD_GAN_models}
\end{figure*}

\subsection{Data preparation}\label{sec:datasets}
We use $r$-band images from the Sloan Digital Sky Survey (SDSS) as a test case though \texttt{PSFGAN} can be applied to any CCD imaging data in any filter. For this proof of concept we choose SDSS data because it is very homogeneous and has many galaxy images available for large training sets. 
We divide the data into a training set, a validation set for model selection, and a testing set to evaluate model performance.  Each set consists of image pairs (original, conditional input). However, only during training \texttt{PSFGAN} uses both the original image without a PS and the conditional input (original image with added PS). In the validation set and the testing set we use exclusively the conditional input as we only run the trained generator on these samples.
To avoid overfitting and ensure the generalization ability of our approach, throughout the whole project, we only use the testing
set \emph{once} for each of the final experiments. The development of models is conducted completely using the validation set.

We test \texttt{PSFGAN} on three redshift ranges corresponding to $z\sim 0.05$, $z\sim 0.1$, and $z\sim 0.2$, respectively.  In these ranges we use $424\times424$ pixels ($168"\times 168"$) cutouts of SDSS galaxies with some variation of redshift to $z\in[0.045, 0.055]$, $z\in[0.095, 0.105]$, and $z\in[0.194, 0.206]$, respectively.  For each redshift sample we split the data into training set of $5000$ images, a validation set of $200$ images and a testing set of $200$ images.

In the following we describe the transformation that we apply to the original images to get the conditional input. We perform the following three steps:
\begin{enumerate}
\item We extract the Point Spread Function (PSF) from the SDSS data.
\item We scale the PSF to a value by a contrast ratio drawn from a predefined distribution.
\item We align the centroid pixel of the galaxy with the centroid pixel of the PSF and then add the images pixel wise.
\end{enumerate}

Hence for a given original image, the corresponding conditional input was defined by two parameters: a) the brightness of the PS and b) the shape of the PSF we convolved it with.

In detail, we implement this procedure differently for the training sets than for the validation and test sets.

\begin{table}
\centering
\begin{tabular}{l|ll}
\hline
 								&  \begin{tabular}{@{}l@{}}\textbf{training set}\end{tabular}   & \begin{tabular}{@{}l@{}}\textbf{testing \&}\\ \textbf{validation sets}\end{tabular}\\
\hline
\textbf{PSF}   				   &  \begin{tabular}{@{}l@{}}analytical fit of \\ SDSS-tool PSF\end{tabular}   & \begin{tabular}{@{}l@{}}$40$-$60$ stars \\combined by \\median stacking\end{tabular}  \\
\hline
\textbf{pdf of $R$} 		  & \begin{tabular}{@{}l@{}} uniform in \\linear space \end{tabular}  & \begin{tabular}{@{}l@{}} uniform in \\logarithmic space\end{tabular}  \\
\hline

\textbf{number of}	  & \begin{tabular}{@{}l@{}} $5000$ \end{tabular} &  \begin{tabular}{@{}l@{}} $200$\end{tabular} \\
\textbf{image pairs}  & &\\
\hline
\end{tabular}
\caption{Overview of the datasets that were used for each of the three redshift groups. $R$ is the contrast ratio which ranged between $0.1$ and $10$ in all cases. To simulate a real application case we introduced a discrepancy between the PSFs in the training and the testing set. The table shows how we simulated the AGN point sources in each distinct case. For a distribution of contrast ratios in the training set we refer to section \ref{sec:linlogtrain}. The size of the datasets was determined heuristically and can be taken as a guideline of what might be appropriate for a general application.}
\label{tab:datasets}
\end{table}

In the training sets, we use the PSF-tool provided by SDSS \citep{sdss} to extract in each image the PSF and fit it with three $2D$-Gaussians in order to get an analytical PSF
\footnote{This tool generates a position dependent, semi-empirical PSF by use of a Karhunen-Lo\`eve transform \citep{sdss}. The fitting step that is performed on the output of the tool is necessary to remove the noise in this PSF image. (The noise would be amplified when the PSF is scaled to high contrast ratios which would lead to unrealistic images.)}.
Its brightness is then scaled by a contrast ratio $R$ defined with respect to the host galaxy luminosity ($cModelFlux\_r$). This contrast ratio is drawn from a uniform distribution in linear space between $0.1$ and $10$, i.\,e. $R\sim U([0.1, 10])$. As we describe in section \ref{sec:GANmodels}, this distribution was chosen because it yields the best performance (among the tested distributions).

In the validation and testing sets, following the approach of \cite{mike}, we measure a semi-empirical PSF by median-stacking $40-60$ stars from the neighborhood of the galaxy. 
(The mismatch between training and testing PSF is necessary to take into account the lack of information about the exact PSF we would have in a real situation.) Due to the high dynamic range of contrasts we want to test for we draw $R$ from a uniform distribution in logarithmic space, i.\,e. $\log(R)\sim U([-1,1])$. 

Table \ref{tab:datasets} shows an overview of the parameters chosen for different datasets.

\subsection{GAN models}\label{sec:GANmodels}
To find a good model we train with different hyperparameters and then evaluate each trained model on the three validation sets. We then choose the model with the overall best performance. We emphasize that we do not perform an exhaustive hyperparameter search. Also the accuracy of a model depends on the random initialization of the weights at the beginning of the training. Therefore, the performance of the selected model represents a lower bound. An exhaustive search for the best hyperparameter and initialization would likely result in superior performance over our limited search.

To quantify the performance we define the recovery ratio as the ratio of recovered PS flux to the real PS flux\footnote{The \emph{real} PS flux is the flux of the PS that we put onto the original image.} and compute its Mean Absolute Deviation (MAD) from $1$ in each validation set. As we will reuse this quantity for various tests in section \ref{sec:results} we simply call it $\Delta$: 
$$
\Delta := \mathrm{average}\left(\left|\frac{\mathrm{recovered\ PS\ flux}}{\mathrm{real\ PS\ flux}}-1\right|\right)
$$
The average is taken over the $200$ instances in the actual validation set. This yields a score for each redshift sample. We average these three scores again in order to obtain a measure for the general accuracy of a model. We then choose the model with the minimal average score.\\
While searching for the best GAN model we vary the following parameters:
\begin{enumerate}
\item preprocessing: normalizing and redistributing pixel values by applying a non linear stretching function
\item distribution of contrast ratios in the training set
\item learning rate (defined in \ref{sec:lr})

\end{enumerate}
Testing the whole parameter space is computationally expensive. Therefore we vary only one parameter at a time while holding the other two parameters at a fixed value. 

We discuss the different models and their scores $\Delta$. Users applying \texttt{PSFGAN} are advised to use our results as a starting point for optimizing the parameters for their specific data.

\subsubsection{Stretch function and scale factor}
It is a common practice to normalize and redistribute the input values of neural networks such that they are comparable across the training set \citep{normalization}. This preprocessing is especially important for this work due to the high contrast between galaxy and PS brightness. If the data was just normalized and scaled linearly, the galaxy would have been interpreted as noise by the GAN in the cases where the PS is very bright. 

Not only the input images themselves have a high dynamic range but also the maximum pixel values across the training set. We want to find a reversible transformation to rescale the images, i.e. redistribute the pixel values in a smaller range. The pixels in the transformed image should be distributed in a way that the GAN is sensible to both the PS and the host galaxy in all of the images. The transformation has to be unique so that it can be applied to all images before showing them to the GAN, and applied back on the output images to recover the full pixel scale.
We test several stretching functions (see \ref{tab:stretchfunc}) while holding the learning rate constant at $\mathrm{lr}=9\cdot 10^{-5}$ and using a uniform distribution in linear space for the contrast ratios in the training set.

We observe (Figure \ref{fig:MAD_GAN_models}) that the $\mathrm{asinh}$ stretch function with a scale factor $A=50$ model has the smallest average $\Delta$.

\subsubsection{Distribution of contrast rations in the training set}\label{sec:linlogtrain}
We test two different distributions of contrast ratios in the training set: a uniform distribution in linear space $R\sim U([0.1,10])$ and a uniform distribution in logarithmic space $\log(R)\sim U([-1,1])$. We hold the stretch function constant at $\mathrm{asinh}, A=50$ and the learning rate at $\mathrm{lr}=9\cdot 10^{-5}$. In Figure \ref{fig:MAD_GAN_models} we plot the scores resulting from evaluation on the validation sets. If \texttt{PSFGAN} is trained on a sample with contrast ratios distributed uniformly in linear space, it is more stable than if it is trained on a sample with contrast ratios distributed uniformly in logarithmic space.

\subsubsection{Learning rate}\label{sec:lr}
The discriminator and the generator are Neural Networks. Therefore they minimize their loss functions by adapting the weights of their neurons. The learning rate determines how much the weights are adjusted in each training step. For a more technical description of the optimization algorithm we are using, we refer to \cite {adam_opt}.

In Figure \ref{fig:MAD_GAN_models} we plot the score $\Delta$ for $6$ different learning rates. While varying the learning rates we hold the stretch function constant at $\mathrm{asinh}$ with a scale factor of $A=50$ and the distribution of contrast ratios in the training set is a uniform distribution in linear space. The model with the lowest average $\Delta$ is the one with $\mathrm{lr}=9\cdot 10^{-5}$.

\begin{table}
\renewcommand{\arraystretch}{2}
\centering
\begin{tabular}{lr} 
\hline
\normalsize\textbf{asinh} & \begin{tabular}{@{}l@{}}\Large$\frac{\text{asinh}(A\cdot x)}{\text{asinh}(A\cdot max)} $ \end{tabular}\\ 
\hline
\normalsize\textbf{log} & \begin{tabular}{@{}l@{}}\Large$ \frac{\log \left( \frac{A\cdot x}{max} \right)}{\log A} $\end{tabular}\\ 
\hline
\normalsize\textbf{pow} & \begin{tabular}{@{}l@{}}\Large$ \sqrt[A]{\frac{x}{max}}  $ \end{tabular}\\ 
\hline
\normalsize\textbf{sigmoid} &\begin{tabular}{@{}l@{}}\Large $2 \cdot \left( \frac{1}{1+e^{\frac{A \cdot x}{max}}}-\frac{1}{2} \right) $\end{tabular}\\
\hline
\end{tabular}
\caption{Overview of the stretch functions used. $A$ is the scaling factor, $max$ refers to the brightest pixel across the whole training set. 
$max = 6140$ for $z \sim 0.05$, 
$max = 1450$ for $z \sim 0.1$,
$max = 1657$ for $z \sim  0.2$}
\label{tab:stretchfunc} 
\end{table}

\begin{figure*}
\centering
\includegraphics[width=\textwidth]{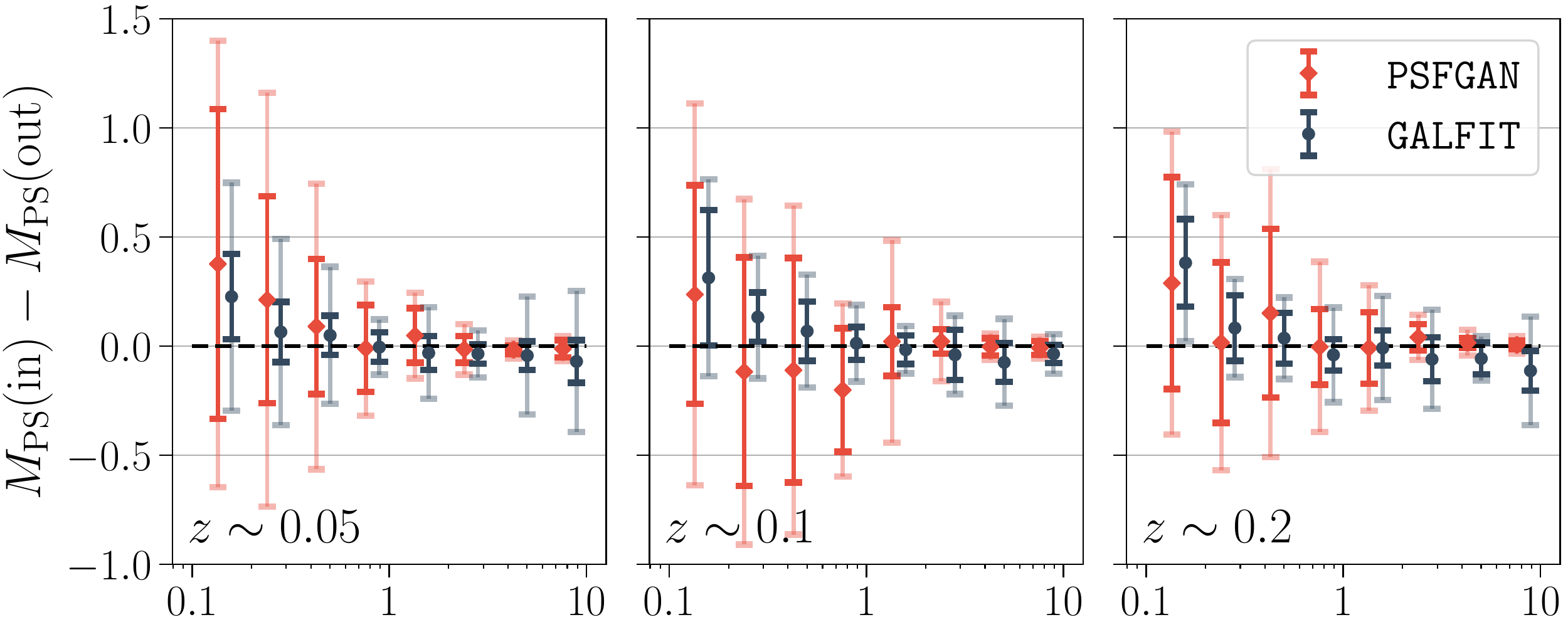}
\includegraphics[width=\textwidth]{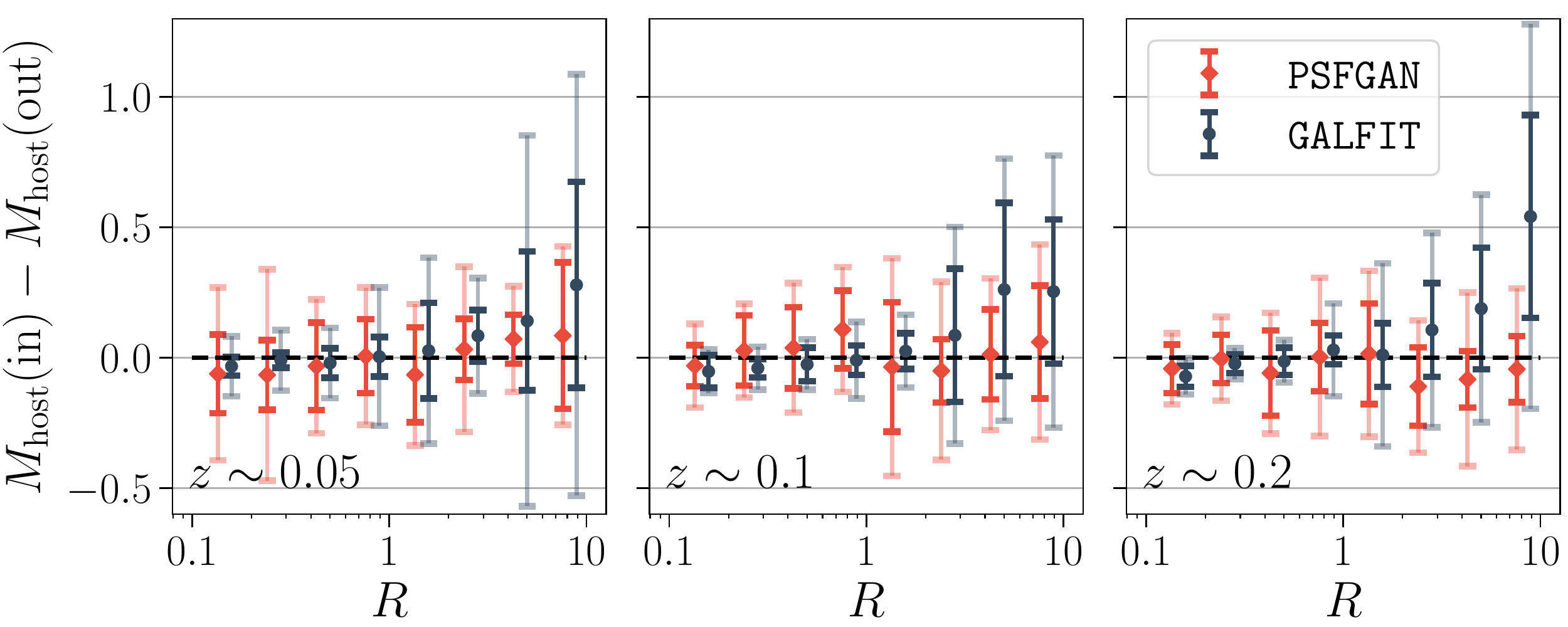}
\caption{We compare how well \texttt{PSFGAN} and \texttt{GALFIT} can recover PS and host galaxy magnitudes by computing the magnitude difference between the input and the output (recovered) for both the PS and the host galaxy. The plotted quantity is the median in its respective bin and the solid (transparent) error bars indicate the distance from the median where at least $68\%$ ($90\%$) of the data points are enclosed. The dotted line indicates perfect recovery ($M_{\mathrm{PS}}(\mathrm{in})=M_{\mathrm{PS}}(out)$ or $M_{\mathrm{host}}(\mathrm{in})=M_{\mathrm{host}}(out)$). At $z\sim 0.05$ we exclude $5$ galaxies from the plot because \texttt{GALFIT} crashed on them. For the same reason we exclude $2$ galaxies from the $z\sim 0.1$ plot and $3$ galaxies from the $z\sim 0.2$ plot.  }
\label{fig:comparisons_PS}
\end{figure*}

\subsubsection{Summary}
Within the subset of the parameter space that we test, we find that the best model is given by the following parameters:
\begin{itemize}
\item learning rate: $\mathrm{lr}=9\cdot 10^{-5}$
\item distribution of contrast ratios in the training set: uniform in linear space
\item preprocessing: $\mathrm{asinh}$ stretch function with a scale factor of $A=50$
\end{itemize}

\begin{figure*}
\centering
\includegraphics[width=\textwidth]{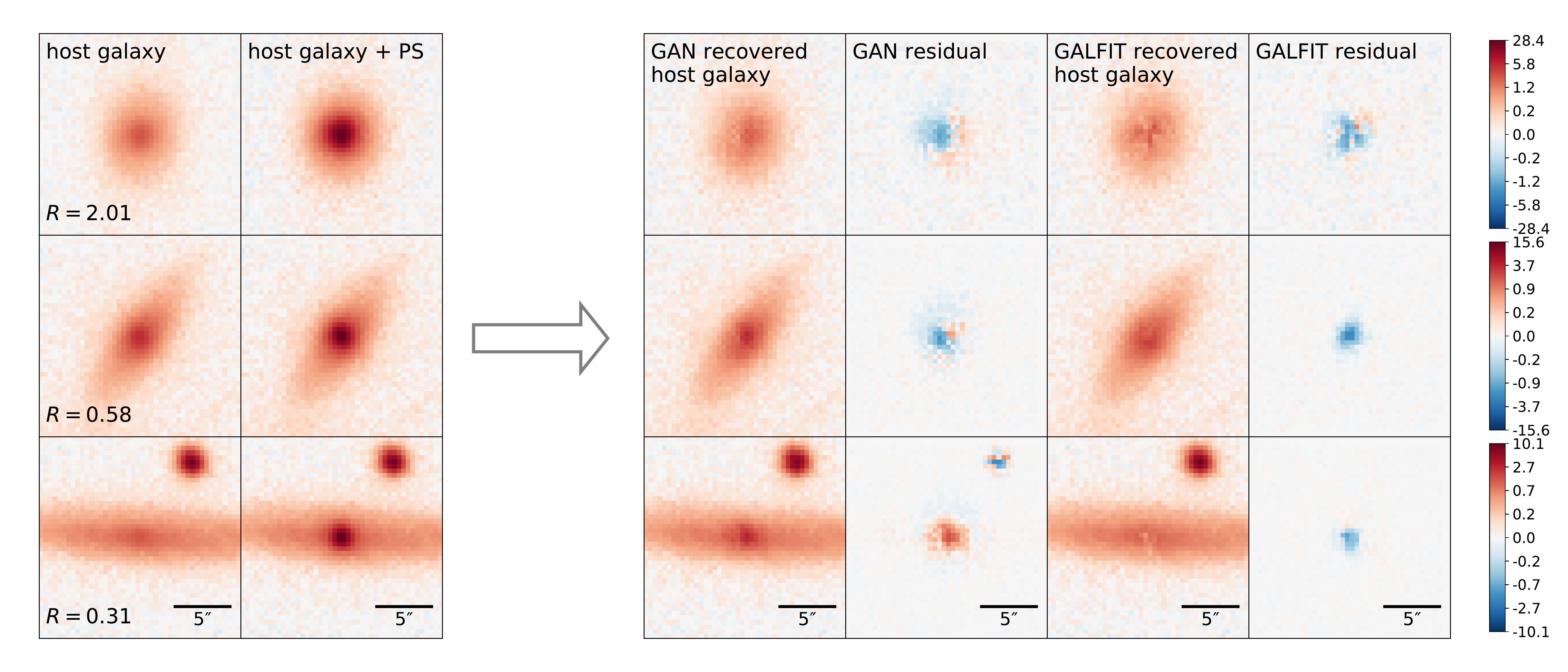}
\caption{Examples at $z\sim 0.05$ with different contrast ratios. In each row we plot the original host galaxy and the host galaxy with the simulated PS in its center. We then plot the output images of \texttt{PSFGAN} and \texttt{GALFIT} to see how they differ from the original galaxy image. We call the output images recovered host galaxy image. Moreover we show resdiuals (recovered-original) for both methods. }
\label{fig:z005ex0}
\end{figure*}
\subsection{\texttt{GALFIT} fitting strategy}\label{sec:fitting_strategy}
In this section we explain the \texttt{GALFIT} fitting strategy we use for the comparisons. \texttt{GALFIT} simultaneously fits an arbitrary number of surface brightness profiles to an image \citep{peng}. Beside various types of inbuilt, analytical function types, it can also fit a PSF provided by the user. A surface brightness component of a specific function type is defined by its geometrical shape and its radial surface brightness profile. For the shape we choose ellipsoids and for the radial surface brightness profile we choose the S\'ersic profile as it is usually done in the literature \citep{schawinksi,mike,simmons}. The S\'ersic profile is defined as:
$$\Sigma_r = \Sigma_e\exp\left(-\kappa\left(\frac{r}{r_e}\right)^{\frac{1}{n}}-1\right)$$
$\Sigma_r$ is the surface brightness at radius $r$, $r_e$ is the half light radius, the S\'ersic index $n$ is a positive real number, $\kappa$ is a parameter which depends\footnote{The parameter $\kappa$ ensures that half of the total flux is always within $r_e$ \citep{peng}.} on $n$ and $r_e$ and the radius $r$ for an ellipsoid is defined by 
$$r = \left(|x|^2+\left|\frac{y}{q}\right|^2\right)^{\frac{1}{2}}$$
where $q$ is the ratio of the minor to major axis of the ellipses describing the isophotes \citep{peng}. To fit the PS component we provide a PSF image as input for \texttt{GALFIT}. We obtain this PSF in the same way as the PSF we use in the training set of \texttt{PSFGAN}: We run SDSS's PSF-tool \citep{sdss} and fit the output with three $2D$-Gaussians. \\
To let \texttt{GALFIT} run in an automated way, we use an approach similar to that of \cite{galapagos}. We run the following algorithm on each galaxy of the testing set:

\begin{enumerate}
\item Run \texttt{GALFIT} with only a PS component to very roughly subtract the PS. This yields an initial guess for the PS flux on the one hand and allows for the next step on the other hand\footnote{If we let \texttt{Source Extractor} run before subtracting the PS, all the host galaxy parameters would be totally biased by the bright PS.}.
\item Run \texttt{Source Extractor} to get initial guesses for host galaxy flux, geometrical parameters and half light radius. 
\item Find stars above the $5\sigma$ limit using the algorithm \texttt{DAOStarFinder} \citep{daofind} and mask them out.
\item Run \texttt{GALFIT} with a S\'ersic component and a PS component. Let the S\'ersic index $n$ be a free parameter within $0$ and $4$. Constrain the magnitude of the host galaxy to be within $\pm 1$ from the initial guess. Moreover restrict the fitting region to a box of $60\,\mathrm{kpc}$ around the galaxy. Leave all the other parameters free.
\end{enumerate}

\begin{figure*}
\centering
\includegraphics[width=\textwidth]{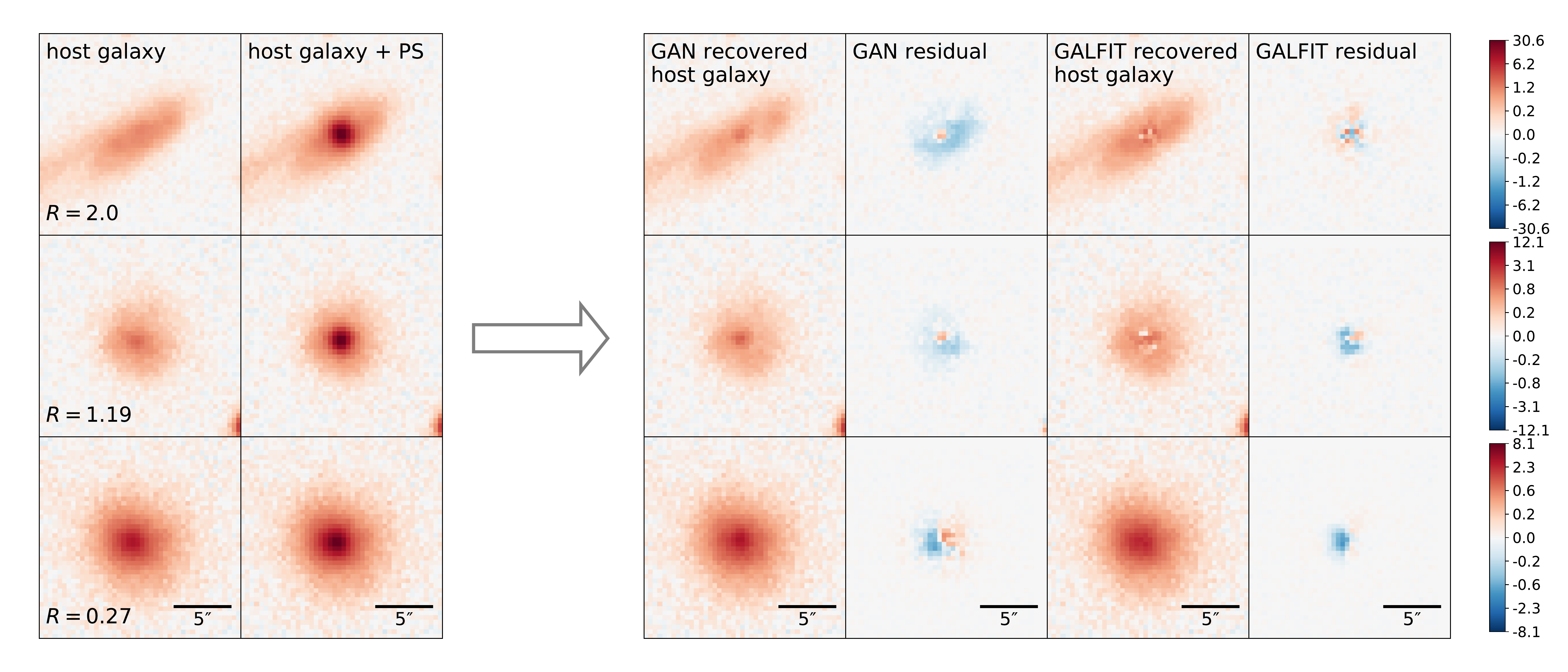}
\caption{Examples at $z\sim 0.1$ with different contrast ratios. The format of the plots is the same than in Figure \ref{fig:z005ex0}.}
\label{fig:z01ex0}
\end{figure*}
\begin{figure*}
\centering
\includegraphics[width=\textwidth]{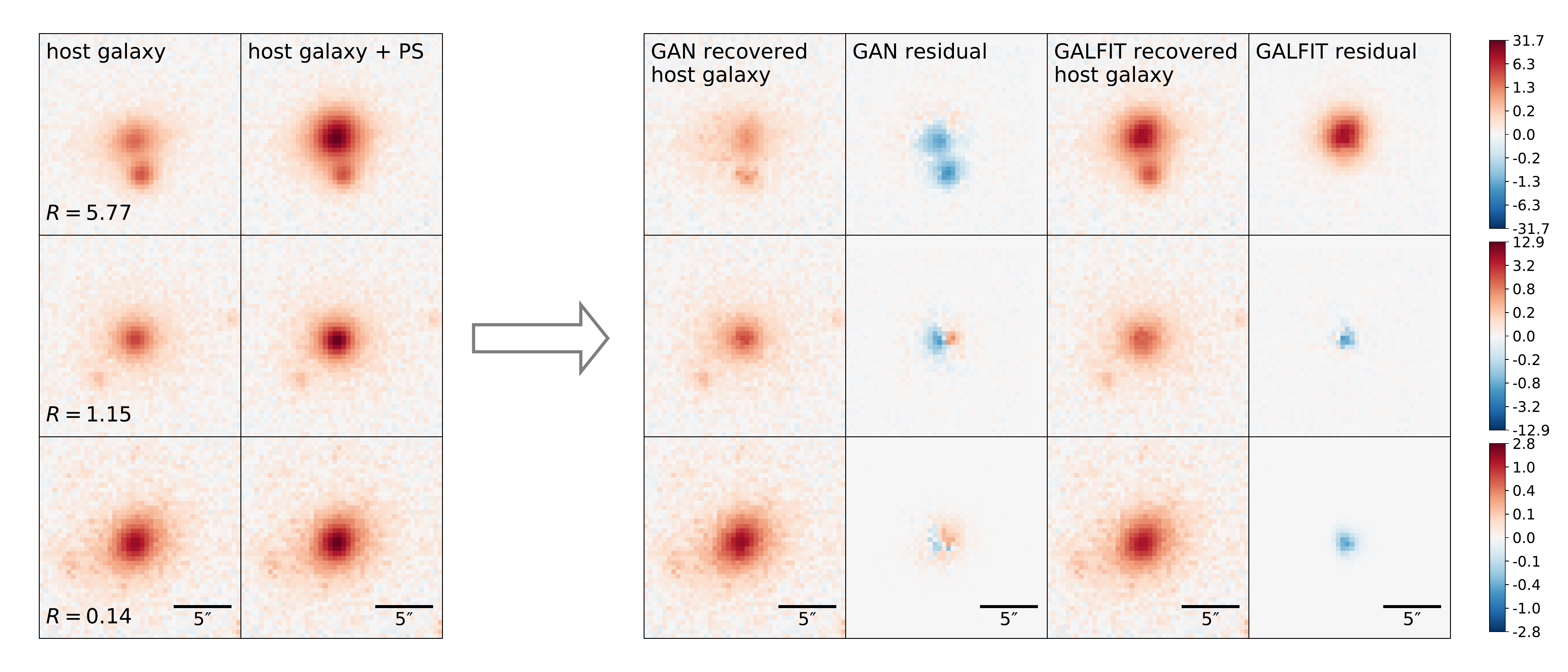}
\caption{Examples at $z\sim 0.2$ with different contrast ratios. The format of the plots is the same than in Figure \ref{fig:z005ex0}.}
\label{fig:z02ex6}
\end{figure*}



\section{Results}\label{sec:results}

We choose the GAN model that works best on the validation set, and evaluate it on the testing sets to produce the results that we present in the following. Section \ref{sec:comparison} contains the comparison to \texttt{GALFIT}. In section \ref{sec:animaltest} we test the dependence of \texttt{PSFGAN} on the brightness distribution underlying to the PS. In section \ref{sec:psf_dependency} we test the sensitivity of \texttt{PSFGAN} on the correct modeling of the PSF and in section \ref{sec:ssi} we test investigate the ability of \texttt{PSFGAN} to recover host galaxy structure. We further test the dependence on the size of the training set in section \ref{sec:tsetsize} and the performance on lower quality data in section \ref{sec:degrading}. Finally, in section \ref{sec:hubble} we explore the behavior of our pretrained models on higher-redshift \textit{Hubble} near infrared data.

\subsection{Comparison of GAN and GALFIT}
\label{sec:comparison}

We quantify the performance by comparing the recovery error in magnitude of both the PS and the host galaxy: We compute the flux of the recovered PS (host galaxy), divide it by the flux of the input PS  (host galaxy) and then convert this ratio to magnitudes. That yields the difference between input PS (host galaxy) magnitude and output PS (host galaxy) magnitude. 

To measure the flux of the recovered PS we subtract the output image (the residual after subtracting the PS component) from the input image and then sum up the pixel values inside a box of $40\times 40$ pixels centered on the center of the galaxy. To measure the flux of the recovered host galaxy we subtract the original image from the output image, sum up the pixel values inside a box of $40\times 40$ pixels centered on the center of the galaxy, and then add the resulting value to the input host galaxy flux which has already been measured by the SDSS pipeline \citep{sdss}. We sum up the pixels using a restricted box because \texttt{PSFGAN} also modifies other sources in the image and we do not want to count those modifications as contributions to the PS flux. As the input host galaxy flux we take the quantity $cModelFlux\_r$ measured by the SDSS pipeline \citep{sdss}. We plot the median magnitude error in different bins of contrast ratios and the $68$ and $90$ percentiles. We define the $\mathrm{n}^{\mathrm{th}}$ percentile as the distance from the median which (in both directions) encloses $n\%$ of the data points.

\begin{figure*}
\centering
\includegraphics[width=\textwidth]{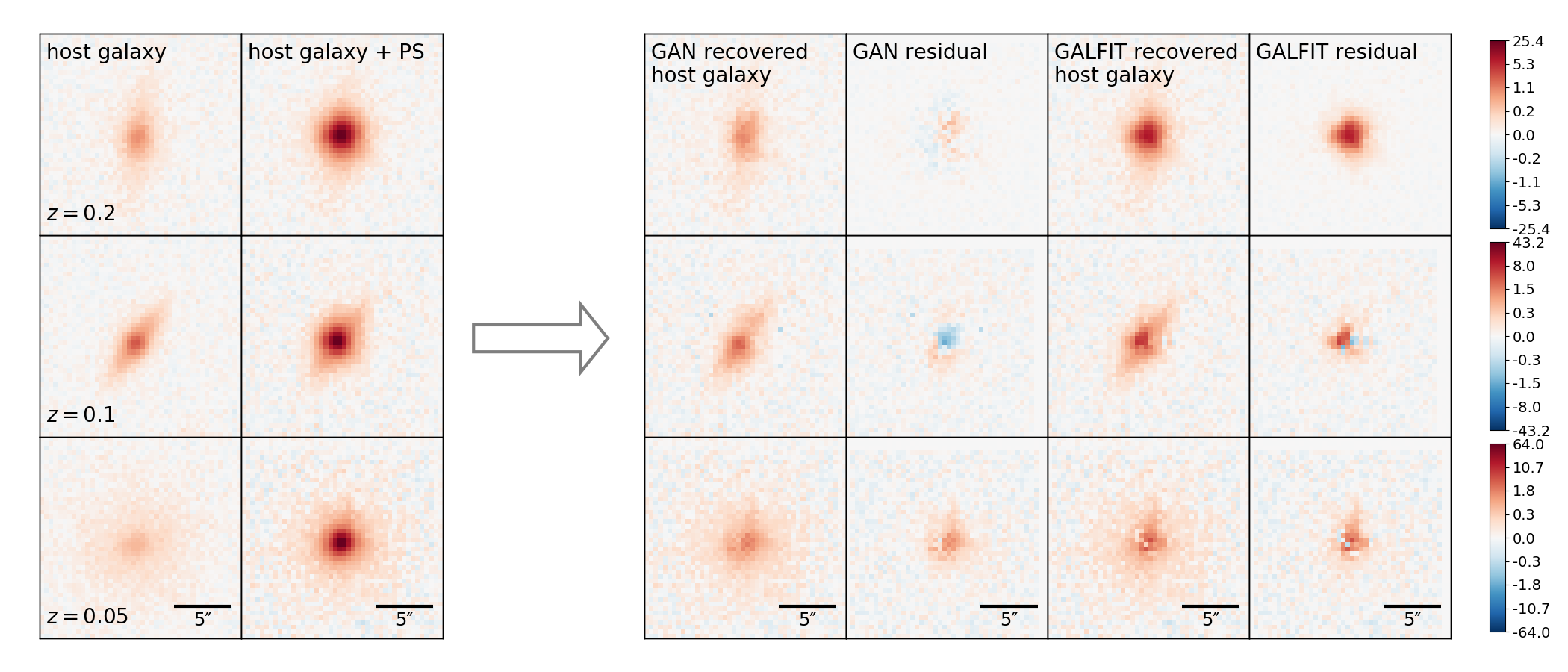}
\caption{High contrast examples in all three redshift samples. The format of the plots is the same than in Figure \ref{fig:z005ex0}.}
\label{fig:ex14}
\end{figure*}

Figure \ref{fig:comparisons_PS} shows the comparison of \texttt{PSFGAN} to \texttt{GALFIT} at the three redshift ranges. Figures \ref{fig:z005ex0}-\ref{fig:ex14} show example images of the original galaxy, the original galaxy with the simulated PS on top of it, the output images (by \texttt{PSFGAN} and \texttt{GALFIT}) as well as residuals (the output subtracted from the original galaxy image). 

Figures \ref{fig:z005ex0} - \ref{fig:z02ex6} show examples of randomly selected contrast ratios in each of the redshift samples. Figure \ref{fig:ex14} shows one high contrast example in each redshift sample.

\begin{table*}
\centering
\begin{tabular}{l|l|l}
\hline
  &\begin{tabular}{@{}l@{}}\texttt{PSFGAN}\end{tabular}   & \begin{tabular}{@{}l@{}}\texttt{GALFIT}\end{tabular}\\
\hline
\textbf{training time} & \begin{tabular}{@{}l@{}} \\takes $\sim 8$ hours for $5000$ images on\\ one NVIDIA Titan Xp GPU \\ \\ \end{tabular} &\begin{tabular}{@{}l@{}} N/A\end{tabular}\\

\hline

\textbf{inference/fitting time} & \begin{tabular}{@{}l}\\
takes $\sim 2\,\mathrm{s}$ per image on a Macbook Air with\\ a $1.7$ GHz  Intel Core i5 CPU and $4$GB RAM \\ (total runtime of $8.3\,\mathrm{h}$ for $500$ images, $\sim 560\,\mathrm{h}$\\ for $10^6$ images)  \\ \\\cline{1-1} \\ takes $\sim 0.15\,\mathrm{s}$ per image on one NVIDIA\\Titan Xp GPU \\ (total runtime of $8\,\mathrm{h}$ for $500$ images, $49.7\,\mathrm{h}$ \\for $10^6$ images) \\\\\end{tabular}
&

\begin{tabular}{@{}l@{}}\\ takes $\sim 7.25\,\mathrm{s}$ per image on the same \\Macbook ($1\,\mathrm{h}$ for $500$ images, $\sim 2000\,\mathrm{h}$\\ for $10^6$ images)\\\\\\
\cline{1-1}\\ not compatible with GPU technology \\\\\\\\\\\end{tabular}

\\
\hline

\textbf{crashes}	  & \begin{tabular}{@{}l@{}} always outputs an image (by construction) \\ \end{tabular} &  \begin{tabular}{@{}l@{}} \\$2.5\%$ crashes for $z\sim 0.05$, $1\%$ crashes for \\$z\sim 0.1$, $1.5\%$ crashes for $z\sim 0.2$\end{tabular}\\& \\
\hline
\end{tabular}

\caption{Comparison of \texttt{PSFGAN} and \texttt{GALFIT} in terms of runtime and robustness. The specified runtime are measured on the $z=0.1$ test sample with images of size $424\times 424$ pixels. \texttt{GALFIT} however only fits a cutout of $83\times 83$ pixels ($\sim 60\,\mathrm{kpc}$). We note that for redshift $z=0.2$ ($z=0.05$) the GALFIT fitting time is smaller (larger) as we fix the fitting region to a box of fixed physical length.}
\label{tab:GALFITproblems}
\end{table*}

Our results show that for contrast ratios $R<1.8$ the median PS magnitude error of \texttt{GALFIT} in general is smaller than that of \texttt{PSFGAN} and reverse for contrast ratios higher than that. For contrast ratios below $R=1.8$ the $68$-percentiles of \texttt{PSFGAN}'s PS magnitude errors are $1.6$-$4.7$ times those of \texttt{GALFIT}. For contrast ratios $R>1.8$ the $68$-percentiles of \texttt{PSFGAN}'s PS magnitude errors are $0.2$-$1.4$ times those of \texttt{GALFIT}. This result is consistent with all redshift samples. For the host galaxy magnitudes we again observe that \texttt{PSFGAN} has smaller systematic error and smaller scatter above $R=1.8$.  For $R<1.8$ the $68$ percentiles of \texttt{PSFGAN} are $1.0-4.6$ times those of \texttt{PSFGAN}. For $R>1.8$ \texttt{PSFGAN} has percentiles smaller than \texttt{GALFIT} with factors between $0.3$ and $1.2$.

In table \ref{tab:GALFITproblems} we compare runtime and robustness of \texttt{PSFGAN} and \texttt{GALFIT}. We find that the fitting time of $\texttt{GALFIT}$ is $\sim 3.6$ times the evaluation time of \texttt{PSFGAN} if they are run on the same machine. By running \texttt{PSFGAN} on GPUs it can be further accelerated such that (in our specific case) it is $\sim 48.3$ times faster than \texttt{GALFIT}. We also find that \texttt{GALFIT} crashes in $\sim 2\%$ of the cases if it is wrapped by our script.

\subsection{Dependence on the underlying brightness profile}
\label{sec:animaltest}

\begin{figure*}
\centering
\includegraphics[width=\textwidth]{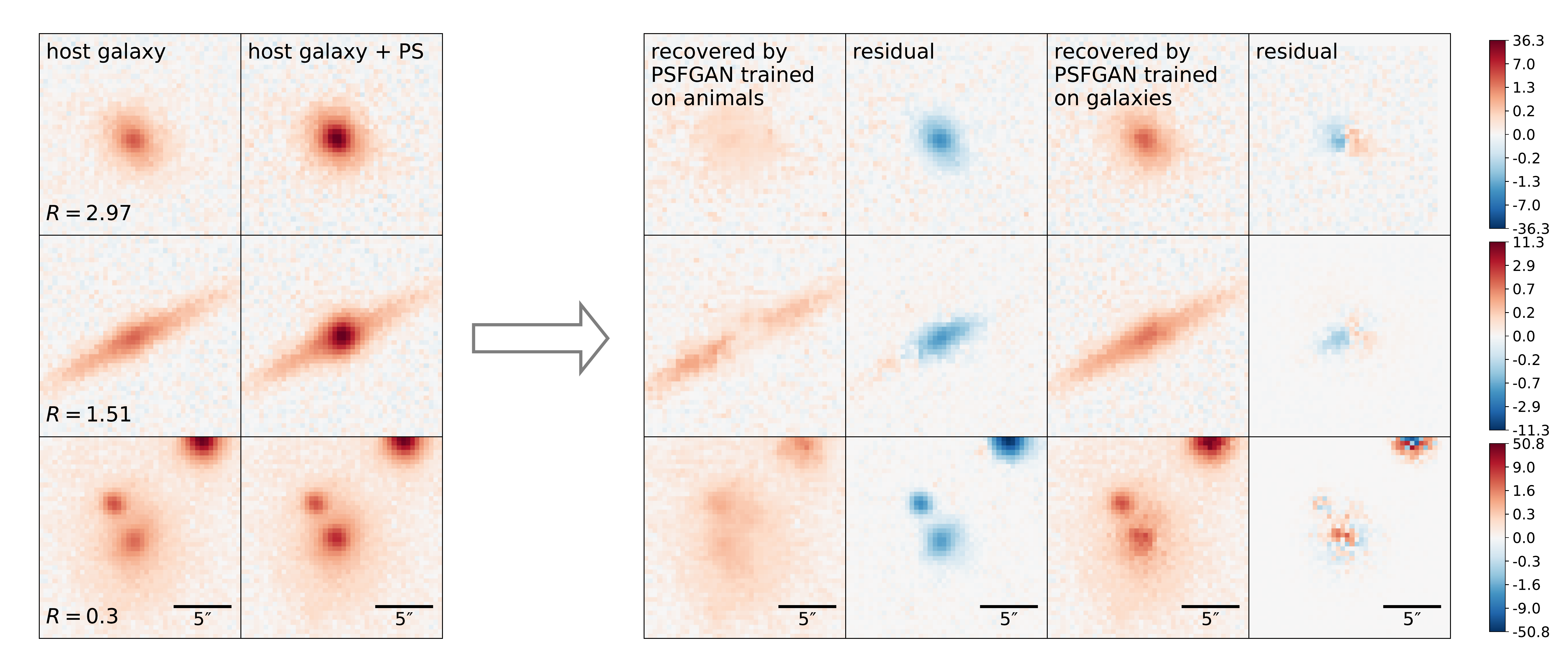}
\includegraphics[width=\textwidth]{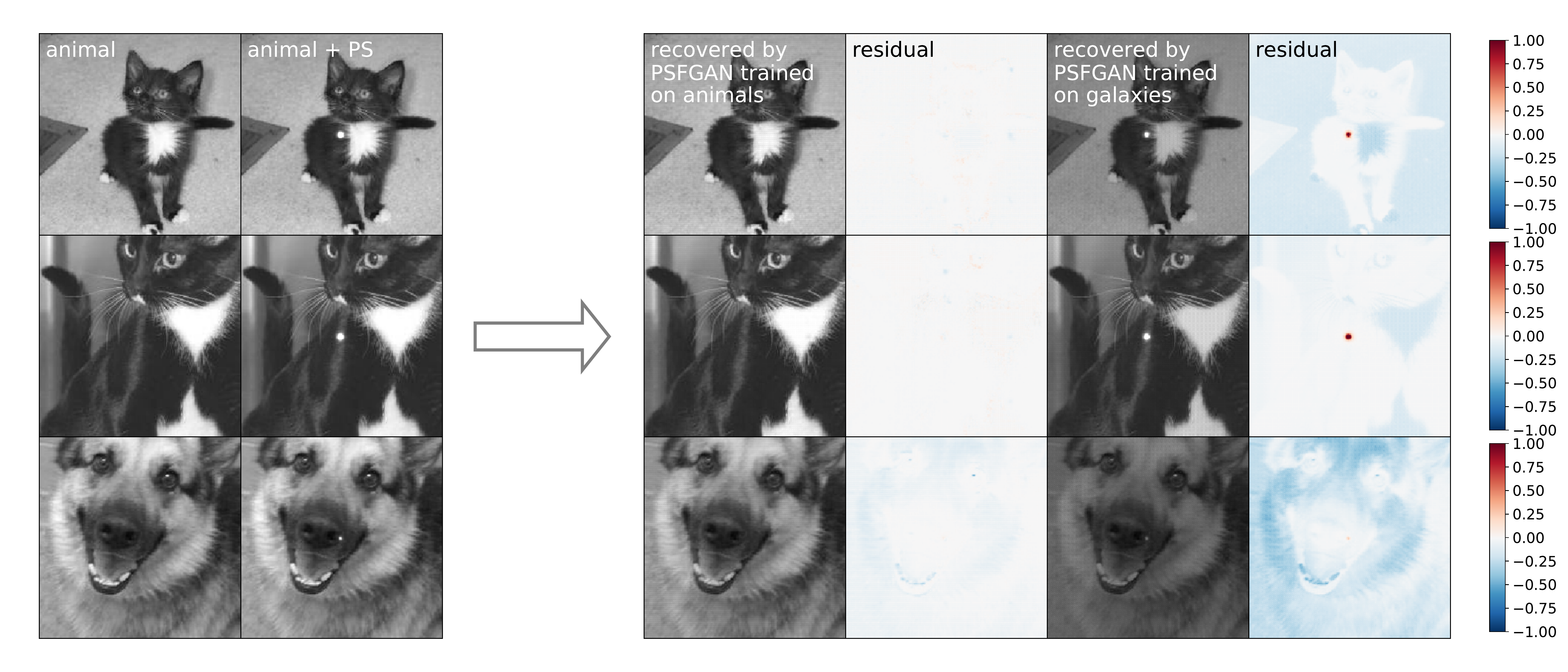}
\caption{We validate the hypothesis
that the visual structure of galaxies helps 
\texttt{PSFGAN} for PS subtraction (i.e., the
neural network learns, intuitively, what galaxies
look like and uses this information for PS subtraction). To validate
this, we apply \texttt{PSFGAN} to very
different domains: cats and dogs. Top: With galaxy images as test images, we compare the outputs and residuals of \texttt{PSFGAN} trained on animals to the ones of \texttt{PSFGAN} trained on galaxies. Bottom: With cats and dogs images as test images, we compare the outputs and residuals of \texttt{PSFGAN} trained on animals to the ones of \texttt{PSFGAN} trained on galaxies. We note that the color map describes only the residuals. We scaled them differently from the animal images in order to visualize both over- and undersubtraction. }
\label{fig:animaltest2}
\end{figure*}

In order to test whether \texttt{PSFGAN} actually uses information of host galaxy brightness distribution we create a comparison sample consisting of pictures of cats and dogs. We add simulated AGN to the centers of the images at different contrasts: We normalize the animal image in such a way that the sum of the pixel values inside a box of $10\times 10$ pixels around the center is equal to the sum of the pixel values inside a box of the same length in the original galaxy image. Although contrast ratio is not well-defined in the case of animals, we plot the PS magnitude recovery against the contrast ratio the PS would have if it was added to the galaxy it corresponds to.\\
We train \texttt{PSFGAN} once on animals and once on galaxies and then evaluate both on each testing set (again one consisting of animals and one consisting of galaxies). 

\begin{figure}
\centering
\includegraphics[width=\columnwidth]{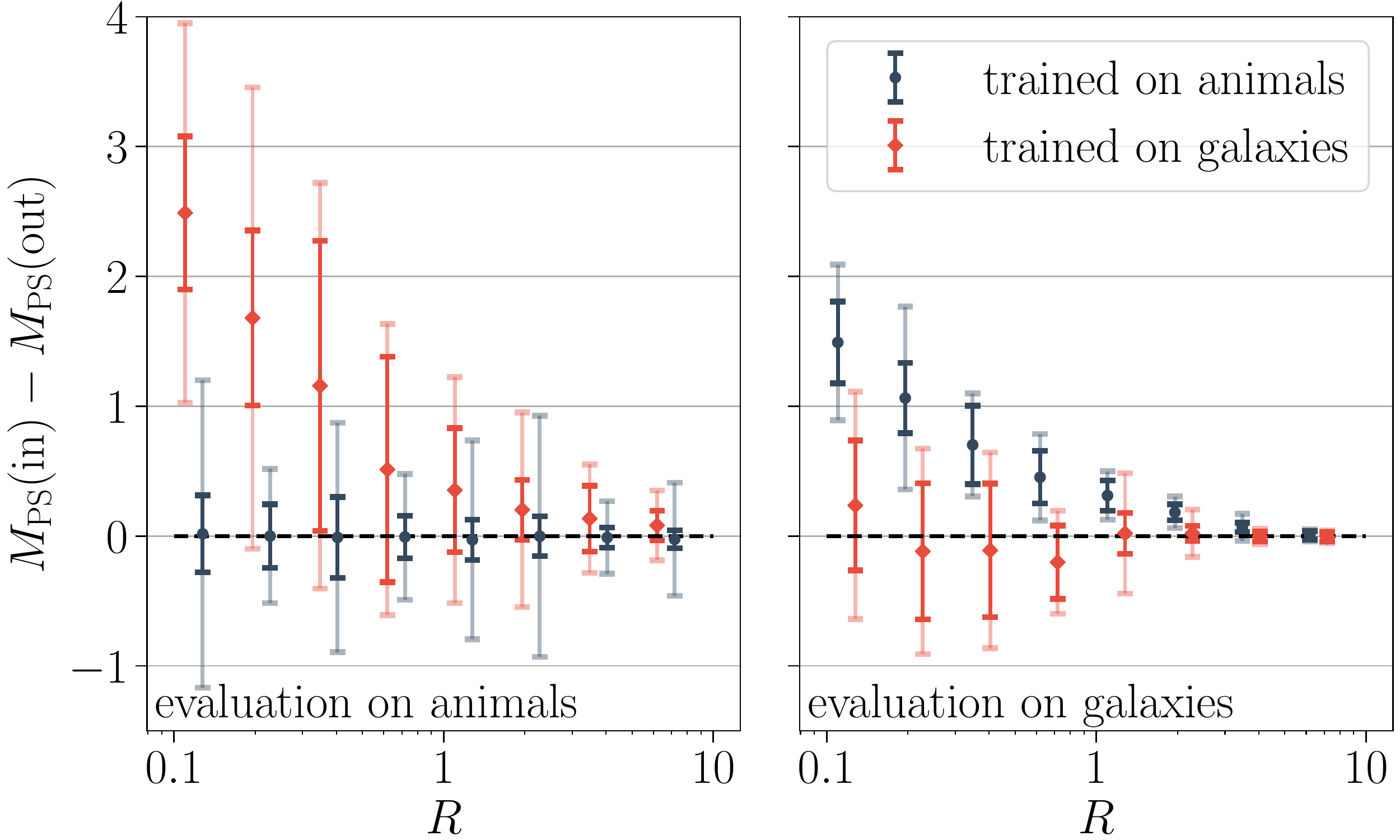}
\caption{A quantitative 
summary of the same experiment
in Figure~\ref{fig:animaltest2}.
The right image shows evaluations on a sample of $200$ galaxies (it is the test sample of $z\sim 0.1$) with artificial point sources and the left image shows evaluations on a set of $200$ animals with artificial point sources.}
\label{fig:animalfluxes}
\end{figure}
Figure \ref{fig:animalfluxes} shows the cross-comparisons and Figure \ref{fig:animaltest2} contains example images. We conclude that the underlying brightness distribution of the objects does indeed matter: \texttt{PSFGAN} trained on animals is better at subtracting point sources from animals and \texttt{PSFGAN} trained on galaxies is better at subtracting point sources from galaxies. However as the contrast increases this effect gets less significant. For evaluation on galaxies both versions of \texttt{PSFGAN} have the same $68$-percentiles in the highest contrast bin $R> 5.6$. Also for evaluation on animals the systematic error and the scatter of \texttt{PSFGAN} gradually decrease with increasing contrast ratios. In the highest contrast bin the version of \texttt{PSFGAN} trained on galaxies has smaller $90$-percentiles. Its $68$-percentile in this bin is twice the $68$-percentile of the version trained on animals.

\subsection{PSF dependence}
\label{sec:psf_dependency}
\begin{table}
\centering
\begin{tabular}{c|c}
\hline
\begin{tabular}{@{}l@{}}\textbf{relative change in}\\\textbf{FWHM of central}\\\textbf{Gaussian}\end{tabular}   & \begin{tabular}{@{}l@{}}\textbf{relative change in}\\\textbf{mean FWHM of}\\ \textbf{single Gaussian fit}\end{tabular}\\
\hline
$\pm 0\%$  & $+0\%$\\

 $+5\%$ & $+2\%$ \\

$-5\%$  & $-2\%$ \\

$+10\%$ & $+5\%$\\

$-10\%$ & $-4\%$\\

$+15\%$ & $+7\%$\\

$-15\%$ & $-6\%$\\

$+20\%$ & $+10\%$\\

$-20\%$ & $-8\%$\\

$+30\%$ & $+15\%$ \\

$-30\%$ & $-11\%$ \\

$+50\%$ & $+27\%$\\

$-50\%$ & $-13\%$\\

$+60\%$ & $+33\%$\\

$-60\%$ & $-21\%$\\

$-70\%$ & $-34\%$\\

$-80\%$ & $-46\%$\\

$-85\%$ & $-52\%$\\

$+100\%$ & $+57\%$\\

$+200\%$ & $+127\%$\\

$+300\%$ & $+200\%$\\

$+500\%$ & $+354\%$\\

\hline
\end{tabular}
\caption{In this table we show how an overall single Gaussian model would be affected by the broadening of the central Gaussian component of our PSF. In the left column we list the relative broadening of the central component of the triple Gaussian PSF. To obtain intuition, we fit the PSF with a single Gaussian before and after broadening it. We then compute the change in FWHM (of the single Gaussian) after broadening relative to the FWHM (of the single Gaussian) before broadening and take the average over the whole set. These are the values in the second column.}
\label{tab:PSFsensitivity}
\end{table}

\begin{figure}
\centering
\includegraphics[width=\columnwidth]{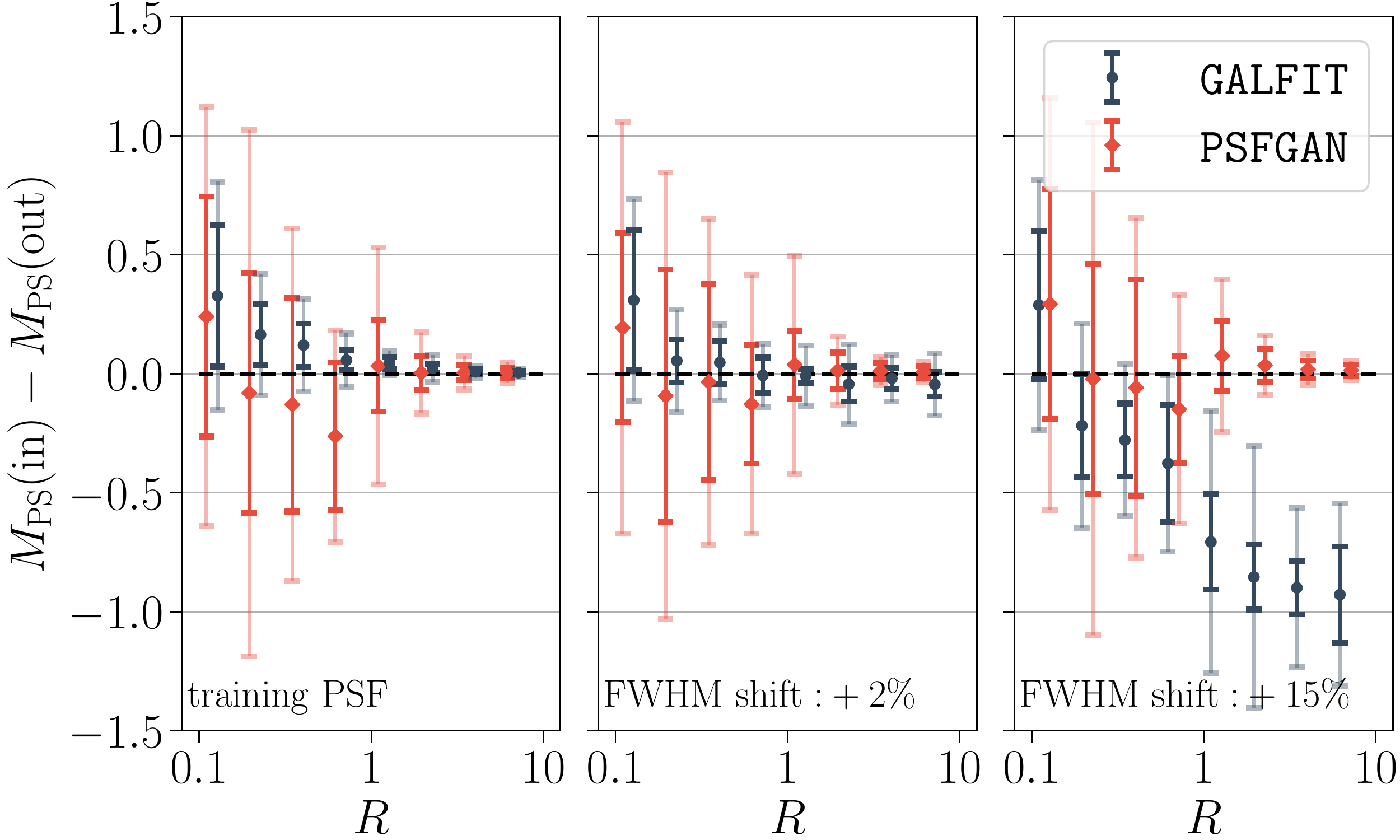}
\caption{Comparison of \texttt{PSFGAN} and \texttt{GALFIT} with increasing broadening of the PSF. We fit the $3$-Gaussian PSF with one single Gaussian before and after broadening its central component and then compute the relative change of the FWHM of this single Gaussian fit. We use this relative change as a measure for the PSF broadening (or narrowing).}
\label{fig:PSFsensitivity_comparison}
\end{figure}

\begin{figure}
\centering
\includegraphics[width=\columnwidth]{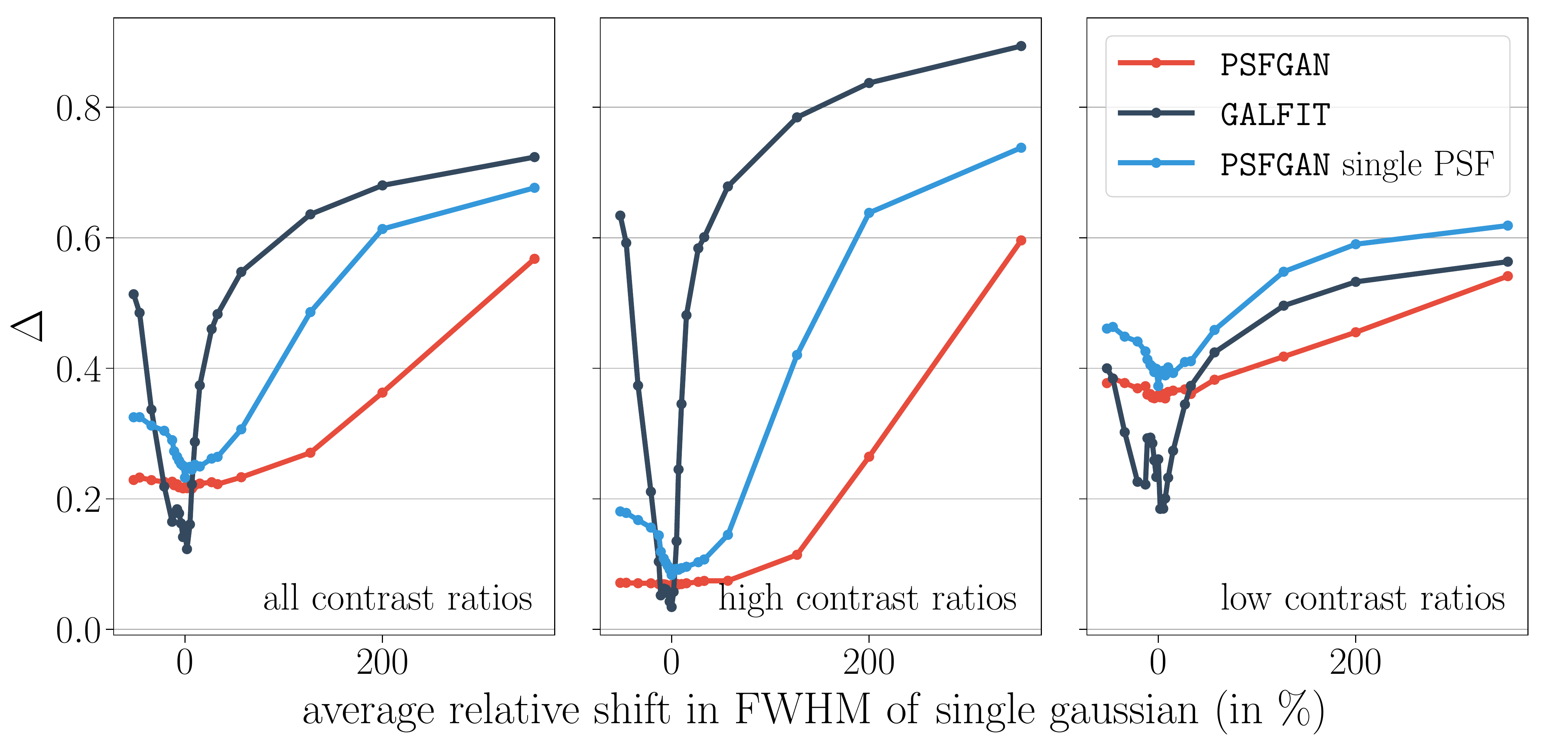}
\caption{Score of \texttt{PSFGAN} and \texttt{GALFIT} for different PSF broadenings. To get the quantity on the $x$-axis we fit the $3$-Gaussian PSF before and after broadening its central component. We then take the relative FWHM change of this single Gaussian fit as a measure for the PSF broadening (or narrowing). The plots show that \texttt{GALFIT} is more sensitive on the PSF than \texttt{PSFGAN}.}
\label{fig:PSFsensitivity_MAD}
\end{figure}

As an example we show a comparison of \texttt{PSFGAN} and \texttt{GALFIT} for broadenings in the single Gaussian FWHM of $+0\%$, $+2\%$ and $+15\%$ in Figure \ref{fig:PSFsensitivity_comparison}. Figure \ref{fig:PSFsensitivity_MAD}
shows the score $\Delta$ for the whole range of FWHM broadenings we tested. We also compare to a version of \texttt{PSFGAN} that was trained on a single PSF. We randomly choose one of the PSF's generated by the SDSS tool and constantly use this one as to simulate the AGN in each galaxy.

The results show that \texttt{GALFIT} has very high accuracy if its input PSF is the same that the one used to simulate the AGN. As soon as there is some discrepancy introduced between those two PSFs \texttt{GALFIT} starts to have large systematic errors. \texttt{PSFGAN} starts to have problems only for broadenings $>100\%$ (in the FWHM of a single Gaussian model). \texttt{PSFGAN} trained on a single PSF is in general (though not for low contrasts) also more robust to PSF variation in the test set. Its $\Delta$ is however higher than that of the normal \texttt{PSFGAN}. Judging from the score $\Delta$, \texttt{GALFIT} can handle a seeing variation of approximately $+ 8\%$ and $-20\%$. However at high contrast ratios $R>1$, \texttt{PSFGAN} has already a lower score for $- 13\%$ and $+5\%$. We conclude that \texttt{PSFGAN} is more robust against seeing variation and improper modeling of the PSF. Moreover we can infer that \texttt{PSFGAN} learns the variation of the PSF during training if it is trained on a variety of PSF's.

\subsection{Host galaxy structure recovery}\label{sec:ssi}
By now we have only tested the recovery of magnitudes. To test how well \texttt{PSFGAN} recovers structure of the host galaxy we use the Structural Similarity index (SSIM). The SSIM is distance metric for two images that takes into account spatial correlations between different pixels \citep{ssim}. The SSIM of two images that are the same is $1$ and it decreases as one of the two images is degraded. As the SSIM was designed to coincide with the quality assessment of the human eye \citep{ssim}, we consider it useful for quantifying the loss and recovery of structural information of AGN host galaxies. For this test we created an additional test sample consisting of spiral galaxies. We compare the structure recovery on this sample to the structure recovery on the normal test samples that we use in this work. It serves as a comparison sample as it consists of mixed types of galaxies. 

To get a sample of spiral galaxies we select galaxies with $z\in [0.04, 0.06]$, $z\in[0.09,0.11]$,  $z\in[0.19,0.21]$ that are neither in the training set nor in the validation set and have Galaxy Zoo vote fractions (for either spiral clockwise or spiral anticlockwise) above $70\%$ \citep{zoo,zoo_dr}. The reason for using a slightly wider redshift range here is that there are not enough sources matching our criteria in the redshift range that we use in the other tests. We finally get a total  of $129$, $168$ and $59$ sources respectively.

\begin{figure*}
\centering
\includegraphics[width=\textwidth]{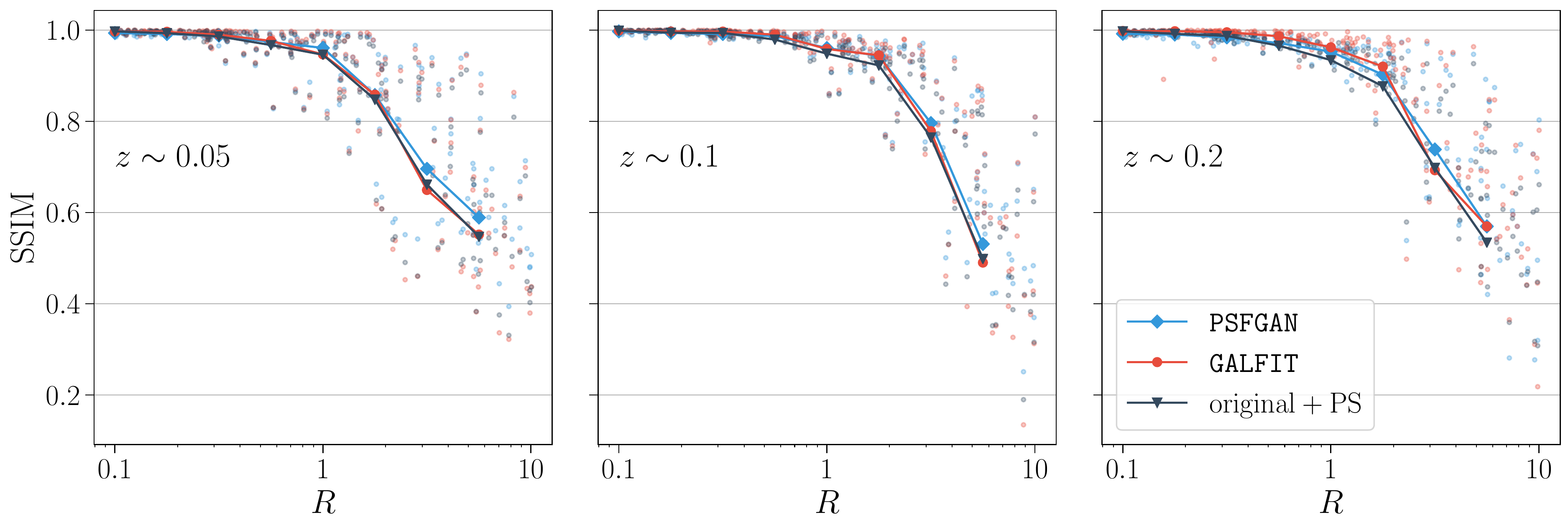}
\caption{SSIM for the mixed type galaxy sample. We observe that \texttt{PSFGAN}'s median SSIM is only higher than that of \texttt{GALFIT} for high contrast ratios.}
\label{fig:ssi_all}
\end{figure*}

In Figure \ref{fig:ssi_all} we compute the SSIM between the original image and the recovered images for both \texttt{GALFIT} and \texttt{PSFGAN}. In order to only extract the relevant information we compute the SSIM on cutouts of the images. We cut out a quadratic box around the center of the galaxy and we chose the length of the box by hand such that the galaxy fills the cutout (therefore we have different box lengths for the different redshift samples). In order to get an intuition for the significance of the different performances we also compute the SSIM between the original galaxy image and the image with the added PS. After plotting each individual SSIM we calculate the median in $8$ bins of contrast ratio and connect the median points with a straight line. 

For the sample of mixed morphologies we find results consistent with the analysis of magnitude recovery. We observe that only above contrast ratio $R\sim 1.8$ \texttt{PSFGAN} has a higher median SSIM than \texttt{GALFIT}. For the spiral galaxies we find that \texttt{PSFGAN} has a higher SSIM already for lower contrast that in the comparison sample of mixed morphologies. We conclude that \texttt{PSFGAN} is less confused by spirals arms.

\begin{figure*}
\centering
\includegraphics[width=\textwidth]{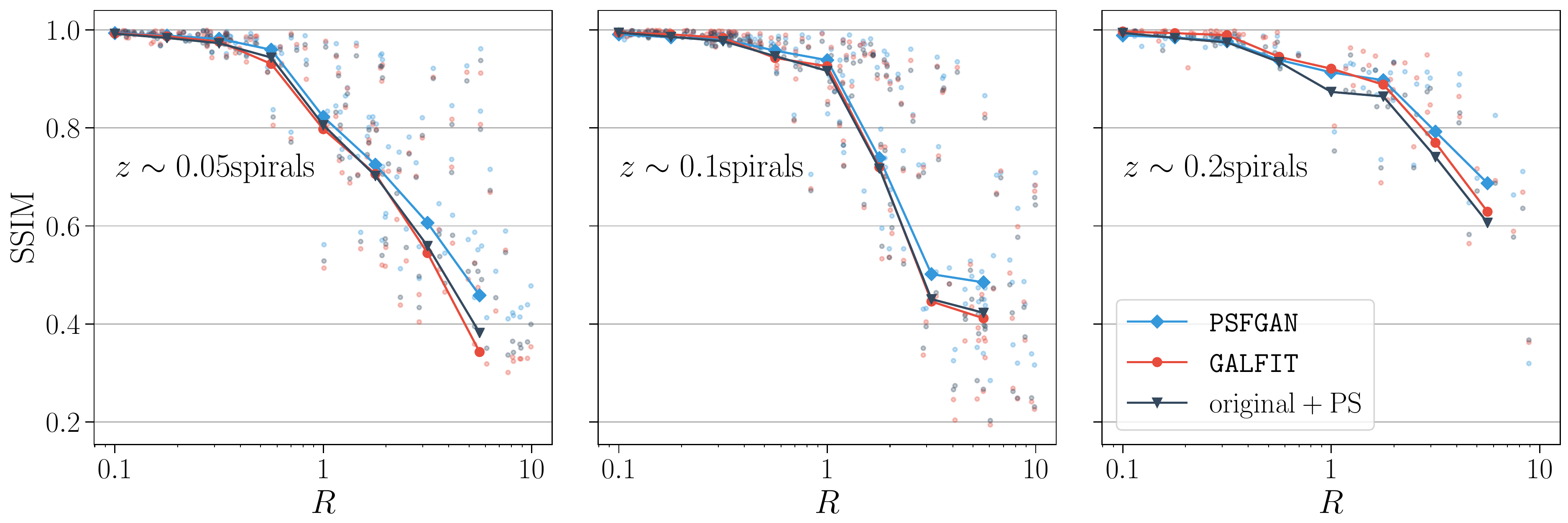}
\caption{SSIM for spirals galaxies. We observe that \texttt{PSFGAN}'s median SSIM is closer to one for moderate and high contrast ratios. For $z\sim 0.05$ and $z\sim 0.1$, \texttt{PSFGAN} has a lower SSIM only for contrast ratios $R<0.6$.}
\label{fig:ssim_spirals}
\end{figure*}

\subsection{Dependence on the size of the training set}\label{sec:tsetsize}
\begin{figure}
\centering
\includegraphics[width=\columnwidth]{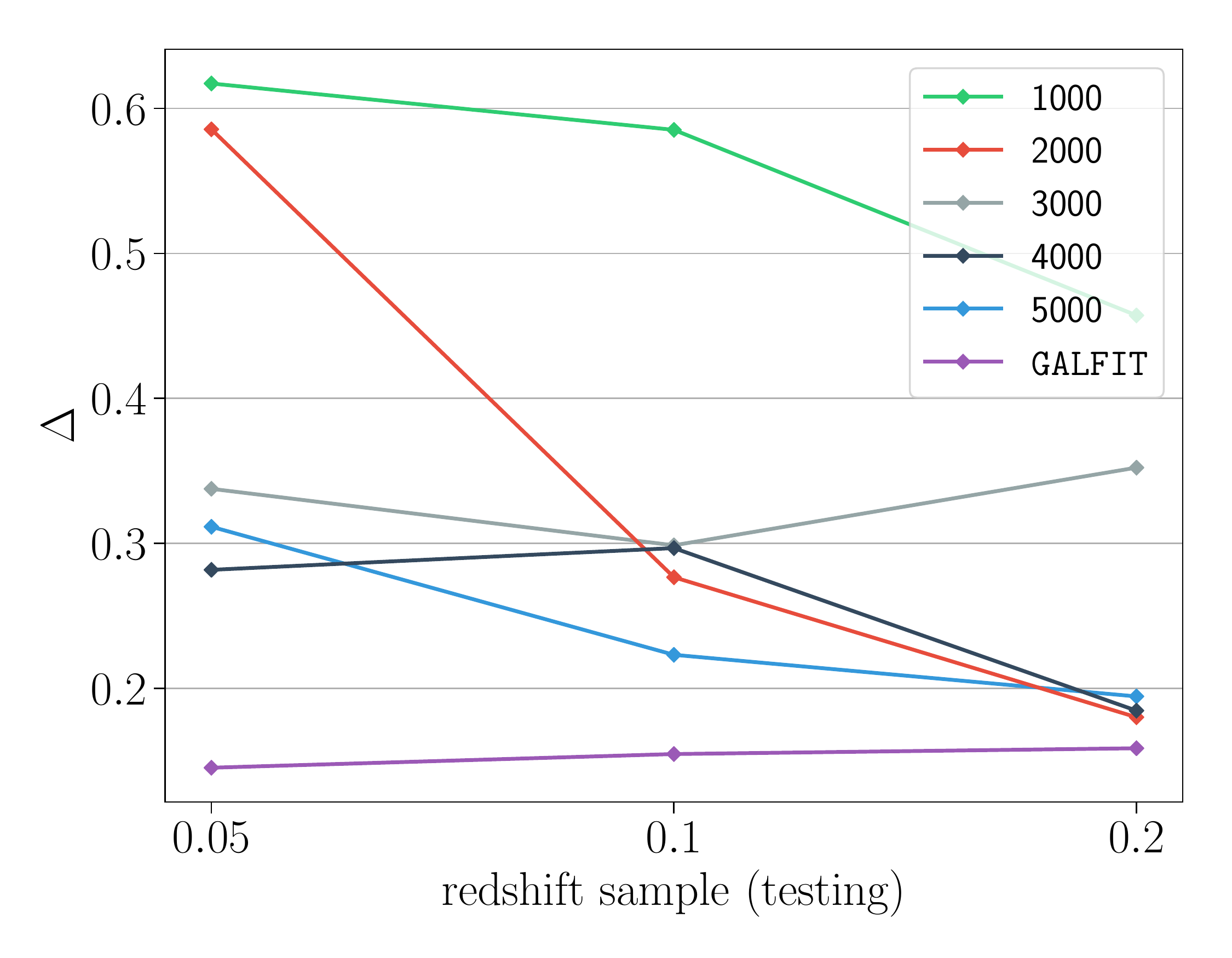}
\caption{We train \texttt{PSFGAN} on training sets with different sizes $1000$, $2000$, $3000$, $4000$ and $5000$. As expected the MAD score increases with decreasing training set size. We conclude that we indeed need approximately $5000$ galaxies to be able to train \texttt{PSFGAN}.}
\label{fig:tsetsize_MAD}
\end{figure}

To test the dependence of \texttt{PSFGAN} on the size of the training set we train (for each redshift) on training sets of size $1000$, $2000$, $3000$, $4000$ and $5000$ images at different redshifts.. We then evaluate on the test samples and compute the MAD of the recovery ratio from $1$ (which we defined as $\Delta$). Figure \ref{fig:tsetsize_MAD} shows how the different models perform. As expected, decreasing the training set leads to a decrease in accuracy.

\subsection{Performance on low quality data}\label{sec:degrading}
The large amount of high quality imaging data provided by SDSS makes it easy to train a GAN. For many applications, the data may be noisier and the resolution poorer. Moreover, finding $5000$ galaxies for the training set is not necessarily feasible for many surveys and wavelengths.

We now show that models trained on SDSS data can perform well on lower quality data. We train a model on degraded SDSS images and compare it to \texttt{GALFIT} and to the model trained on non-degraded images. We compare the models by evaluating them on a degraded test sample. To degrade the images we convolve the original image with a Gaussian kernel of size $5\times 5$ pixels and FWHM $FWHM_{\mathrm{kernel}}$. We then add white noise with a variance such that the noise variance of the degraded image $\sigma_d$ is larger than the initial noise variance $\sigma_i$ of the original image. For each redshift we create three differently degraded tests with $(\mathrm{FWHM}_{\mathrm{kernel}}, \sigma_d/\sigma_i)=(1.2, 1.5),(1.2, 1.8),(2.0, 1.8)$. The way we degrade the training PSF is different from the way we degrade the PSF in the test sets. For the test sets we convolve the PSF image obtained by median combining stars with the same kernel and add the resulting image to the degraded galaxy image. We do not add white noise to the PSF image as we already add noise to the whole original image.

In the training set we degrade the PSF image obtained from the SDSS tool by applying the same transformation as for the original images. Then we fit the degraded PSF image with two Gaussians. Fitting three Gaussians is not possible here because the convolution smoothes out the images.)

To compare the models we again use the one-dimensional score $\Delta$ from section \ref{sec:GANmodels}. We estimate the performance of the model trained on non-degraded images as well as the model trained on degraded images by evaluating them on a degraded test set. We then run \texttt{GALFIT} on the degraded test set where we provide a degraded PSF image as input. To get the input PSF we perform the same steps than for creating the degraded PSF in the training sets. We apply Gaussian blurring and add white noise to the image that is outputted by the SDSS PSF tool and then fit the resulting image with two Gaussians. We choose the variance of the white noise such that the noise of the PSF image gets increased by a factor $\sigma_d/\sigma_i$.

\begin{figure*}
\centering
\includegraphics[width=\textwidth]{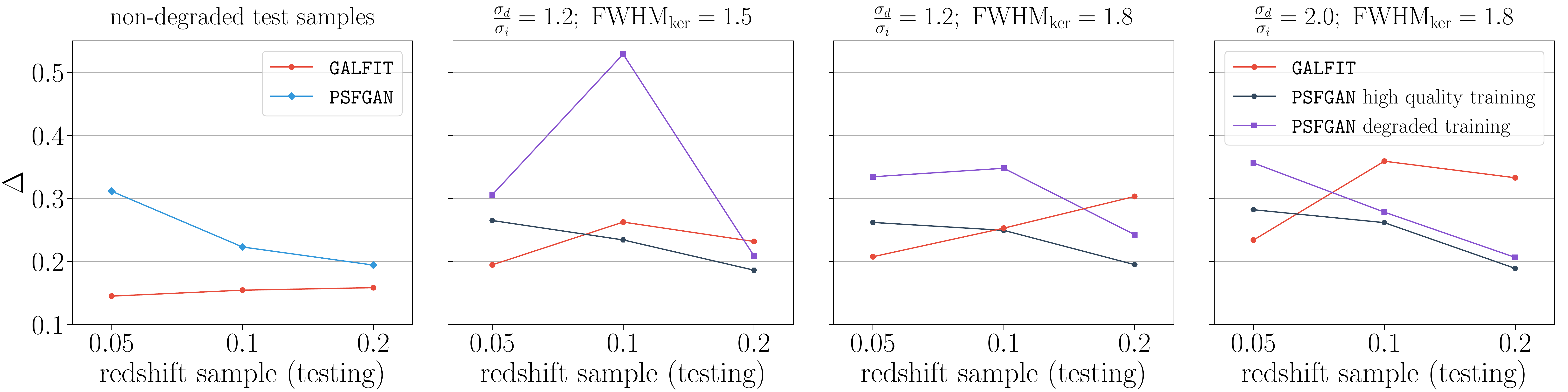}
\caption{We degrade the test sample of each redshift by applying convolution with a Gaussian kernel and adding white noise. We create three different degraded test sets according to different kernel FWHMs and white noise variances. The quantity $\sigma_d/\sigma_i$ is the noise variance of the degraded image divided by the noise variance of the original image.}
\label{fig:MAD_degrading}
\end{figure*}

Figure \ref{fig:MAD_degrading} shows the scores for the three degraded test sets and compares them to the performance of \texttt{PSFGAN} and \texttt{GALFIT} on the non-degraded test set. The plots show that both \texttt{GALFIT} and \texttt{PSFGAN} have a larger $\Delta$ if they are run on the degraded test samples. However \texttt{PSFGAN} is more stable. For the non-degraded images \texttt{GALFIT} has a lower score for all three redshifts. For the most strongly degraded images ($\sigma_d/\sigma_i=2.0$, $\mathrm{FWHM}_{\mathrm{ker}}=1.8$) \texttt{GALFIT} only has a lower score for redshift $z=0.05$. For the other two samples both \texttt{PSFGAN} models have a lower score.

\subsection{Applying \texttt{PSFGAN} to \textit{Hubble} data}\label{sec:hubble}
\begin{figure*}
\centering
\includegraphics[width=\textwidth]{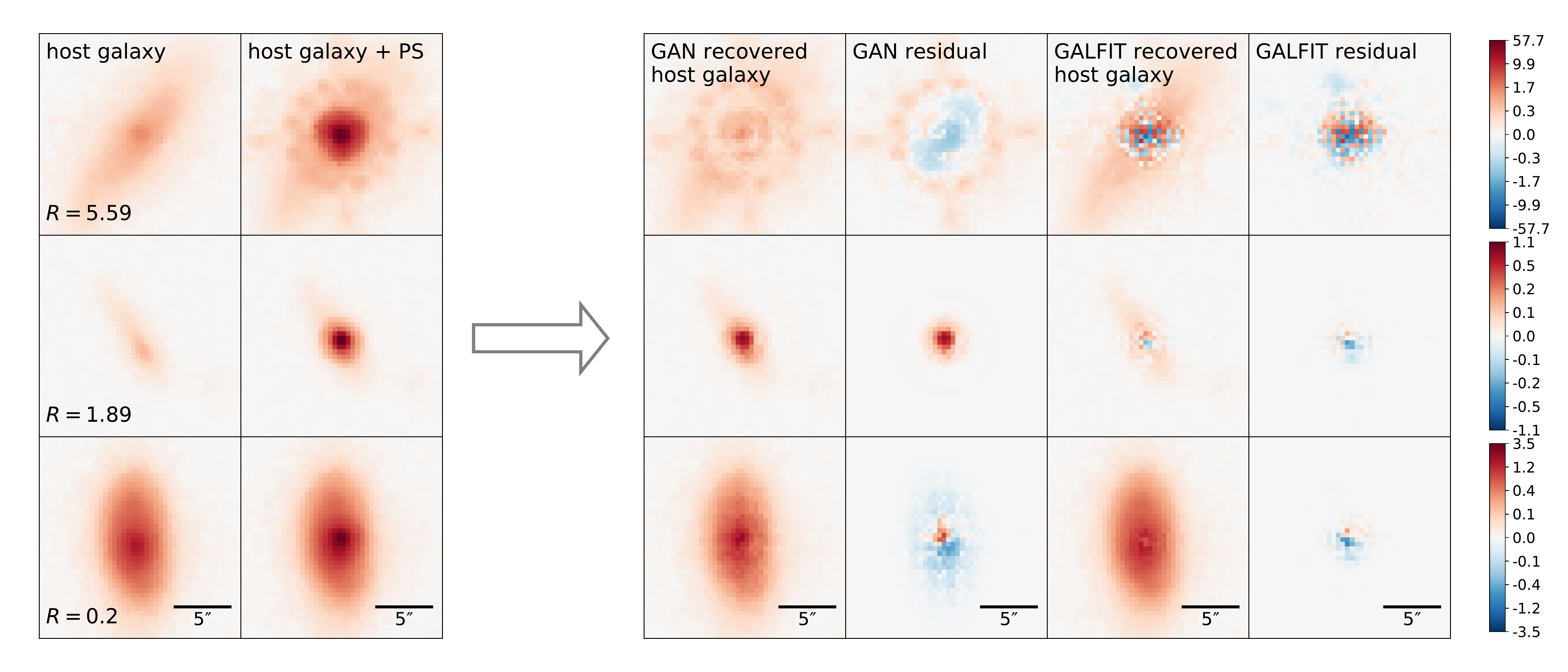}
\caption{Examples of \textit{Hubble} WFC3 images in the F160W filter of galaxies with $z\in[0.4, 0.5]$. We randomly choose three examples with different contrast ratios. The format of the plots is the same than in Figure \ref{fig:z005ex0}.}
\label{fig:ex15}
\end{figure*}

\begin{figure*}
\centering
\includegraphics[width=\textwidth]{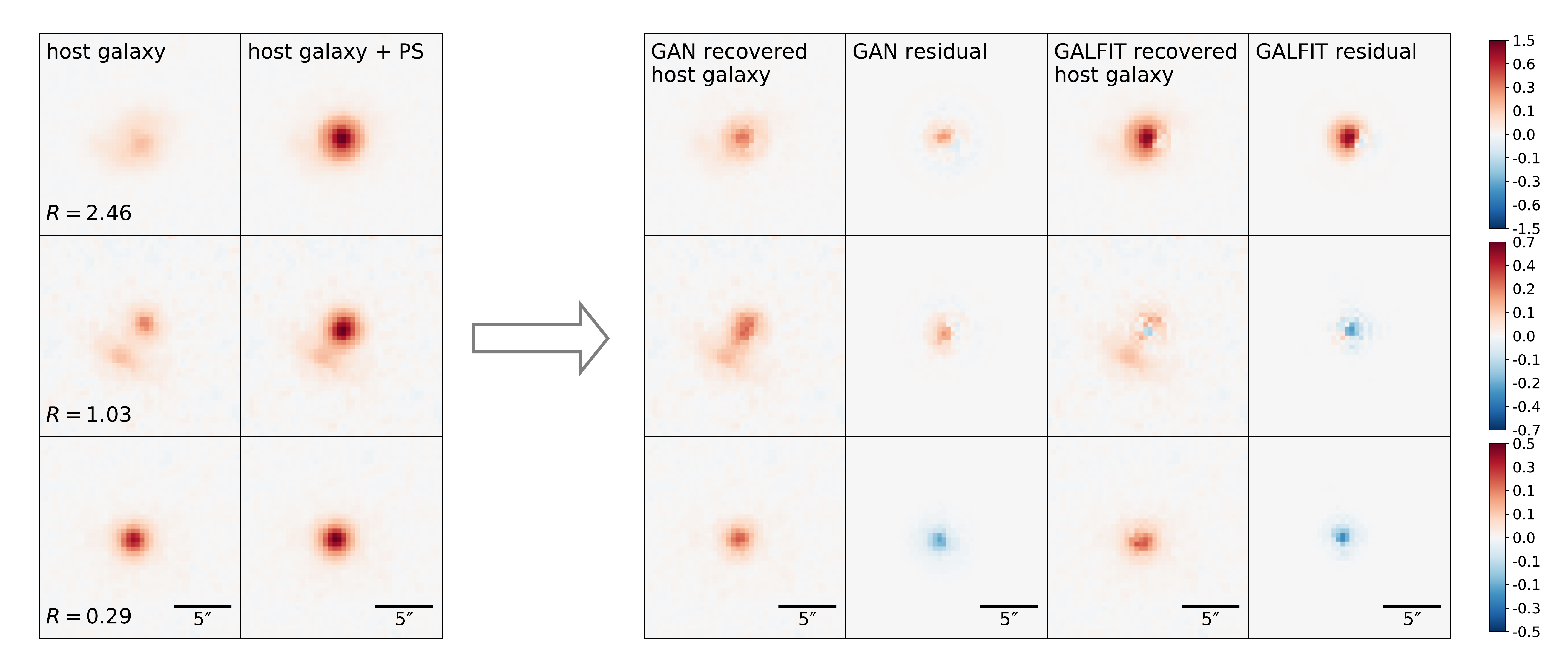}
\caption{Examples of \textit{Hubble} WFC3 images in the F160W filter of galaxies with $z\in[1.0, 1.5]$. We randomly choose three examples with different contrast ratios. The format of the plots is the same than in Figure \ref{fig:z005ex0}.}
\label{fig:ex2}
\end{figure*}

To demonstrate that \texttt{PSFGAN} can be used even if there is not enough training data available, we apply it to GOODS-S WFC3 data in the F160W filter \citep{candles_1,candles_2}. We use the fully calibrated, drizzled images. We create two test sets with different redshift ranges. We use the GOODS-S CANDELS stellar mass catalog \citep{santini} to select detections with\texttt{Source Extractor}'s flag "Star\_Class" $< 0.8$ and observed AB magnitude in the F160W filter $m<24$. We exclude detections with "AGN\_Flag"$<0.1$. This yields a set consisting of $164$ galaxies with $z\in[0.4, 0.5]$ and another set consisting of $195$ galaxies with $z\in[1.0, 1.5]$. We simulate the AGN point sources by stacking $10-30$ stars from the neighborhood of the galaxy. We combine the stacked stars by taking the weighted median in each pixel where we distribute the weights according to the signal-to-noise ratio.

We then evaluate our pretrained \texttt{PSFGAN} models and compare them to the \texttt{GALFIT} script we described in section \ref{sec:fitting_strategy}. The PSF image we provide as input for \texttt{GALFIT} is a cutout of the brightest star with $S/N>100$ we can find in the whole field. In Figure \ref{fig:ex15} and \ref{fig:ex2} we show example images of the original host galaxy, the host galaxy with the PS in its center, \texttt{PSFGAN} and \texttt{GALFIT} recovered host galaxies as well as both method's residuals. The examples show that \texttt{PSFGAN} is not able to subtract the extended wings of the \textit{Hubble} PSF which is intuitive given the fact that it was trained on the SDSS PSF. 

\begin{figure*}
\centering
\includegraphics[width=\textwidth]{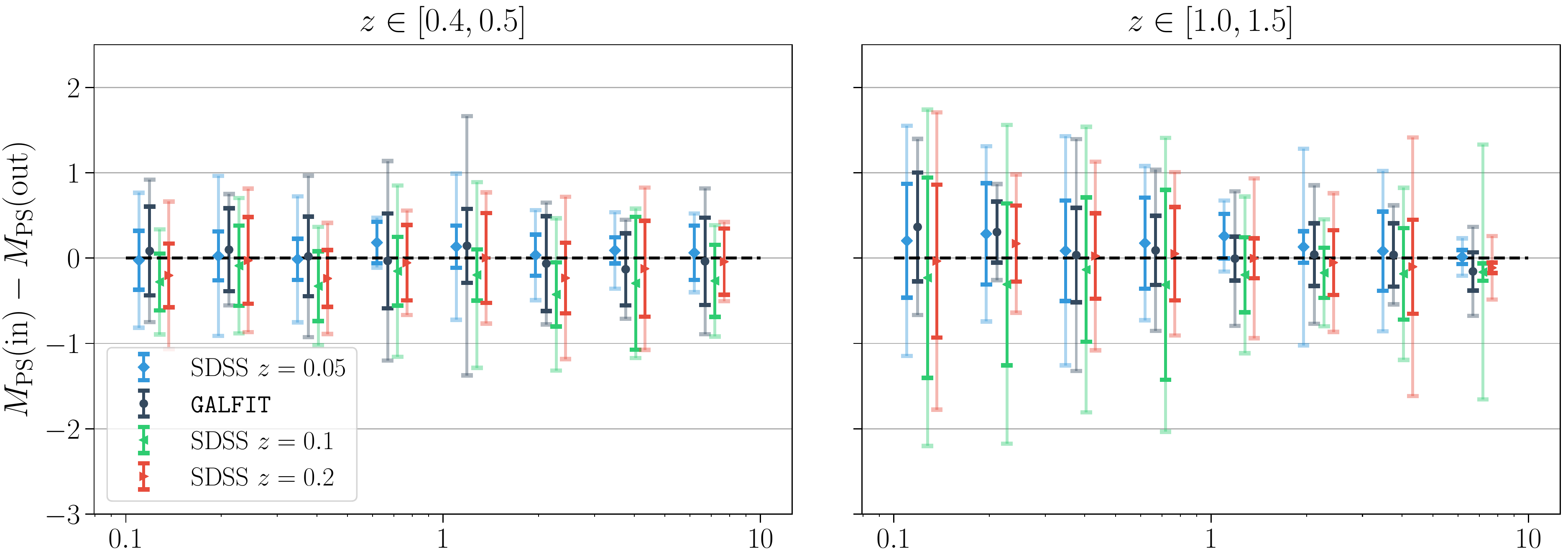}
\includegraphics[width=\textwidth]{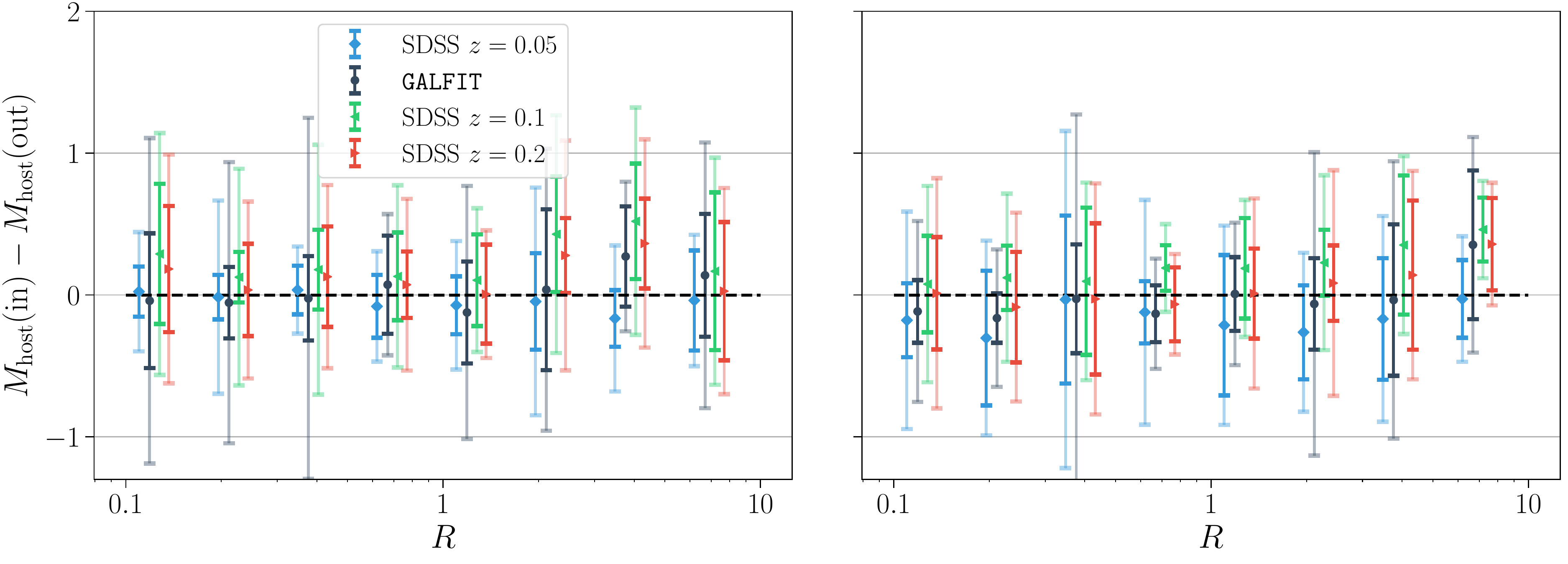}
\caption{PS magnitude and host magnitude errors of evaluation on \textit{Hubble} WFC3 galaxies in two different redshift samples (left plots and right plots). The plots show that the \texttt{PSFGAN} models trained on SDSS data (with the SDSS PSF) have similar bias and variance than the \texttt{GALFIT} script we used.}
\label{fig:hubble_comparisons}
\end{figure*}

Figure \ref{fig:hubble_comparisons} shows the PS magnitude errors and the host magnitude errors for the test sample with $z\in[0.4, 0.5]$ and the test sample with $z\in[1.0, 1.5]$. 
For all models we exclude $18$ galaxies from the left plots and $30$ from the right plots because \texttt{GALFIT} crashed on them. To compute the medians and the percentiles we only use those galaxies where the recovered flux is positive for both the PS and the host galaxy for all of the models. For some cases ($ < 10\%$) there is another source within the restricted box we use to compute the recovered PS flux. In the case where \texttt{PSFGAN} increases the brightness of this close source the computation of the recovered PS flux can result in a negative flux value ($<2\%$). The recovered host galaxy flux can be negative if either \texttt{PSFGAN} or \texttt{GALFIT} massively over-subtracts ($<3\% for \texttt{GALFIT}$ and no observed cases for \texttt{PSFGAN}).
All in all we have to exclude another $4$ galaxies for $z\in[0.4, 0.5]$ and $6$ for $z\in[1.0,1.5]$.

The tests show that the models pretrained on SDSS data can indeed be applied to different data and they even have accuracy comparable to \texttt{GALFIT}. Evaluated on the test sample with $z\in [0.4,0.5]$, the $z\sim 0.05$ SDSS model has a smaller scatter and similar systematic error than our \texttt{GALFIT} script. At $z\in [1.0, 1.5]$ it is difficult to read a significant difference by eye just from the plots of the magnitude errors. Therefore we also list the $\Delta$ scores in table \ref{tab:MAD_hubble}. At first we notice that the score is higher for all models than if they are evaluated on SDSS data. The $\Delta$ is increased by a factor of $1.1$ for the $z\sim 0.05$ model, by a factor of $1.7$ for the $z\sim 0.1$ model and by a factor of $2.0$ for the $z\sim 0.2$ model. Evaluation on the high redshift test sample yields an increase by factor $2.1$,$2.2$,$2.8$ for the respective model of redshift $z\sim 0.05, 0.1, 0.2$. However \texttt{GALFIT}'s score increases as well and a thorough comparison reveals that all the SDSS models have a lower score than our \texttt{GALFIT} script for both test samples.

\begin{table}
\renewcommand{\arraystretch}{2}
\centering
\begin{tabular}{l|cc}
\hline
&$z\in [0.4,0.5]$&$z\in[1.0, 1.5]$ \\
\hline
\texttt{GALFIT} &$0.51$&$0.68$ \\
\hline
$z\sim 0.05$ SDSS model & $0.34$&$0.64$ \\
\hline
$z\sim 0.1$ SDSS model	  & $0.37$&  $0.49$\\
\hline
$z\sim 0.2$ SDSS model  &$0.38$ &$0.55$\\
\hline
\end{tabular}
\caption{For both \textit{Hubble} WFC3 test sets we compute the $\Delta$ score for \texttt{GALFIT} and the SDSS models trained on the three redshift samples. (Before computing $\Delta$ we exclude all the galaxies where \texttt{GALFIT} crashed in the results of the \texttt{PSFGAN}.) We find that all pretrained models have a lower $\Delta$ for both test samples.}
\label{tab:MAD_hubble}
\end{table}




\section{Discussion}\label{sec:discussion}
We have shown that GANs can be used to make photometric measurements. \texttt{PSFGAN} is able to separate AGN point sources from their host galaxies. We have shown that \texttt{PSFGAN} intuitively learns the light distribution of galaxies and applies this knowledge to subtract the PS. For contrast ratios above $R=1.8$ it recovers PS and host galaxy fluxes with a smaller median magnitude error and a lower scatter than a single S\'ersic + PS fit performed by \texttt{GALFIT}. We observe that for low contrast ratios ($R<1.8$) \texttt{PSFGAN}'s scatter in PS magnitude recovery is $1.6$-$4.7$ times larger than \texttt{GALFIT}'s scatter and for high contrast ratios ($R>1.8$ ) \texttt{GALFIT}'s scatter is up to $5$ times the scatter of \texttt{PSFGAN}. We have found that - in terms of SSIM - \texttt{PSFGAN} can recover host galaxy structure of spiral galaxies at least as good as a single S\'ersic + PS fit performed by \texttt{GALFIT} while being better with higher contrast ratios. For $z\sim 0.05$ and $z\sim 0.1$ \texttt{PSFGAN} has a higher median SSIM already for $R=0.3$. To conclude that \texttt{PSFGAN} can handle complicated morphologies better than parametric fitting in batch-mode further tests should be conducted. 

Parametric fitting is very powerful for well-resolved galaxies and low contrast ratios. However it struggles at high contrast ratios $R>1.8$ because of the degeneracy between PS magnitude and host magnitude. Indeed, in this contrast range, \texttt{GALFIT} artificially increases the S\'ersic index which causes the PS to be underestimated. This behavior is documented in the literature \citep{kim,mike}.

The fact that \texttt{PSFGAN} performs well at high contrast ratios makes it a promising tool for studying AGN and their host galaxies at higher redshift where classical methods tend to break down. Indeed with increasing redshift the contrast ratio tends to be higher as the intrinsic emission emerges from a bluer part of the Spectral Energy Distribution (SED) where the AGN is dominant. Also the host galaxy is affected by surface brightness dimming while the PS is not \citep{dimming}. This again increases the probability of finding high contrast systems with increasing redshift. 

We have shown that \texttt{PSFGAN} is more stable with noisier and lower resolution imaging data. Evaluated on differently degraded data we find that \texttt{GALFIT} always has a lower $\Delta$ than \texttt{PSFGAN} for $z\sim 0.05$. However for $z\sim 0.1$ and $z\sim 0.2$ the accuracy of \texttt{GALFIT} declines faster (with the decline in quality) than \texttt{PSFGAN}'s accuracy. For a kernel width $\mathrm{FWHM}_{\textbf{ker}}=1.8$ and noise variance $\sigma_d= 2.0*\sigma_i$ the $\Delta$ score of \texttt{GALFIT} increases by more than a factor $2$ compared to the evaluation on non-degraded images. The $\Delta$ of \texttt{PSFGAN} (trained on high-quality data) increases by a factor less than $1.2$. Furthermore we find that \texttt{PSFGAN} trained on non-degraded images has a lower $\Delta$ on degraded-images than if it was trained on degraded images. We conclude that it can better learn the light distribution of galaxies if the training data is of high quality.

We find that it is indeed necessary to have a training set size of $\sim 5000$ images. However if not enough data is available and a training can not be performed, the user can also apply the \texttt{PSFGAN} trained on SDSS data. We demonstrate that our pretrained models can be applied on \textit{\textit{Hubble}} IR data up to redshift $z=1.5$. Although the accuracy is lower on this data than it was on SDSS data, it compares well to our \texttt{GALFIT} script. For the \textit{Hubble} test sample with $z\in [0.4,0.5]$ the best model is the one trained on $z\sim 0.05$ SDSS data. Its $\Delta$ score is $67\%$ of that of\texttt{GALFIT}. For the \textit{Hubble} test sample with $z\in [1.0,1.5]$ the best model is the one trained on $z\sim 0.1$ SDSS data with a score of $72\%$ of that of \texttt{GALFIT}. We find that in agreement with section \ref{sec:animaltest} the $z\sim 0.05$ SDSS model performs best on the more nearby sample and the $z\sim 0.1$ SDSS model performs best on the more distant higher redshift sample.

The inference phase of \texttt{PSFGAN} is faster on a CPU and one can accelerate it further by running it on GPUs. Run on a Macbook Air with a $1.7$
GHz Intel Core i5 CPU and $4$GB RAM it is $\sim 3.6$ times faster than
\texttt{GALFIT} run on the same machine. By running
\texttt{PSFGAN} on GPUs it can be accelerated such that its inference phase is more than $\sim 40$ times faster than GALFIT \footnote{These numbers should serve as rough estimation as they are specific for our implementation and hardware.}. The strength of \texttt{PSFGN} however lies in its ability to apply the same trained model to many images. If a low number of galaxies is considered \texttt{GALFIT} may have a speed advantage due to \texttt{PSFGAN}'s training time of $\sim 8$ hours (on a GPU). However, e.g. for $10^6$ galaxies the total runtime of \texttt{PSFGAN} (training + evaluation) is only $2.5\%$ of \texttt{GALFIT}'s runtime.

The lack of input parameters during evaluation is another strong advantage of \texttt{PSFGAN}. Unlike parametric fitting methods which are very sensitive on their input parameters, \texttt{PSFGAN} is very robust and requires no human interaction once it is trained. Also it requires fewer physical assumptions than parametric fitting. The only physical knowledge that goes into \texttt{PSFGGAN} is the training PSF. A user has to model the PSF of the data to simulate the point sources in the training set. We have however found that for using \texttt{PSFGAN} it is less important to correctly model the PSF than for using \texttt{GALFIT}. \texttt{PSFGAN} is thus especially powerful to analyze ground based data where the seeing is variable. 

Although we have trained \texttt{PSFGAN} to subtract AGN point sources in SDSS data, it is neither limited to AGN nor to SDSS data. \texttt{PSFGAN} is a general framework for subtracting point sources from CCD images in an automated way. In order to apply \texttt{PSFGAN} to some specific case a user should go through the following procedure:
\begin{enumerate}
\item Create a training set consisting of pairs of images (original, original+point source). We used real observations of galaxies but if there is not enough data available a user could also simulate the ground truth (e.\,g. use simulated galaxies). If ground based images are used, make sure \texttt{PSFGAN} sees a variety of PSFs during the training.
\item Look at the histogram of pixel values of the whole training set. Then decide which stretching function might be appropriate. We recommend starting with $asinh$ and trying different scale factors.
\item Test the setup on a separate testing set to estimate the accuracy. 
\end{enumerate}

We propose a number of applications of our method. One task, that \texttt{PSFGAN} may be suited for is subtraction of fore-ground stars from galaxy images. The only difference from subtracting quasar point sources is the position of the point source relative to the galaxy. Another task where \texttt{PSFGAN} could be applied to is separating supernovae from their host galaxies. Given that this is usually done by fitting galaxy templates, \texttt{PSFGAN} could both simplify and accelerate those measurement processes. Lastly we propose to apply our method to quasar spectra. Like images of quasar host galaxies, their spectra are as well contaminated the AGN. Indeed the architecture of \texttt{PSFGAN} can easily be adapted for taking spectra as input. However the training process might be less straight-forward than in our case where the quasar was a point source and thus had a (more or less) constant shape. 

The code of \texttt{PSFGAN} is described at \href{http://space.ml/proj/PSFGAN}{http://space.ml/proj/PSFGAN} and available at \href{https://github.com/SpaceML/PSFGAN/}{https://github.com/SpaceML/PSFGAN/}. Moreover we will provide the pretrained models at $z\sim 0.05$, $z\sim 0.1$ and $z\sim 0.2$.

\section*{Acknowledgements}

KS, LFS, and AKW acknowledge support from Swiss National Science Foundation Grants PP00P2\_138979 and PP00P2\_166159, and KS from the ETH Zurich Department of Physics.  CZ and the DS3Lab gratefully acknowledge the support from the Swiss National Science Foundation NRP 75 407540\_167266, IBM Zurich, Mercedes-Benz Research \& Development North America, Oracle Labs, Swisscom, Zurich Insurance, Chinese Scholarship Council, the Department of Computer Science at ETH Zurich, and the cloud computation resources from Microsoft Azure for Research award program.  MK acknowledges support from NASA through ADAP award NNH16CT03C and the Swiss National Science Foundation through the Ambizione fellowship grant PZ00P2 154799/1.  Funding for the SDSS and SDSS-II has been provided by the Alfred P. Sloan Foundation, the Participating Institutions, the National Science Foundation, the U.S. Department of Energy, the National Aeronautics and Space Administration, the Japanese Monbukagakusho, the Max Planck Society, and the Higher Education Funding Council for England. The SDSS Web Site is http://www.sdss.org/.  The SDSS is managed by the Astrophysical Research Consortium for the Participating Institutions. The Participating Institutions are the American Museum of Natural History, Astrophysical Institute Potsdam, University of Basel, University of Cambridge, Case Western Reserve University, University of Chicago, Drexel University, Fermilab, the Institute for Advanced Study, the Japan Participation Group, Johns Hopkins University, the Joint Institute for Nuclear Astrophysics, the Kavli Institute for Particle Astrophysics and Cosmology, the Korean Scientist Group, the Chinese Academy of Sciences (LAMOST), Los Alamos National Laboratory, the Max-Planck-Institute for Astronomy (MPIA), the Max-Planck-Institute for Astrophysics (MPA), New Mexico State University, Ohio State University, University of Pittsburgh, University of Portsmouth, Princeton University, the United States Naval Observatory, and the University of Washington. Finally this work is based on observations taken by the CANDELS Multi-Cycle Treasury Program with the NASA/ESA HST, which is operated by the Association of Universities for Research in Astronomy, Inc., under NASA contract NAS5-26555.




\bibliographystyle{mnras}
\bibliography{citations.bib} 

\bsp	
\label{lastpage}
\end{document}